\documentclass[10pt,pra,aps,showpacs,twocolumn,superscriptaddress]{revtex4-1} 
\usepackage{lipsum}
\usepackage{graphicx}
\usepackage{amssymb}
\usepackage{amsmath}
\usepackage[usenames, dvipsnames]{color}
\usepackage[normalem]{ulem}
\usepackage{lipsum}
\begin{document}
\date{\today}
\title{Miscibility in coupled dipolar and non-dipolar Bose-Einstein condensates}
\author{Ramavarmaraja Kishor Kumar}
\affiliation{Instituto de F\'{i}sica, Universidade de S\~{a}o Paulo, 05508-090 S\~{a}o Paulo, Brazil}
\author{Paulsamy Muruganandam}
\affiliation{Department of Physics, Bharathidasan University, Tiruchirappalli 620024, Tamilnadu, India}
\author{Lauro Tomio}
\affiliation{Instituto de F\'isica Te\'orica, Universidade Estadual Paulista, 01156-970 S\~ao Paulo, SP, Brazil}
\affiliation{Instituto Tecnol\'ogico de Aeron\'autica, DCTA,12.228-900 S\~ao Jos\'e dos Campos, SP, Brazil.}
\author{Arnaldo Gammal}
\affiliation{Instituto de F\'{i}sica, Universidade de S\~{a}o Paulo, 05508-090 S\~{a}o Paulo, Brazil}
\begin{abstract}
We perform a full three-dimensional study on miscible-immiscible conditions for coupled dipolar and 
non-dipolar Bose-Einstein condensates (BEC), confined within anisotropic traps. Without loosing general 
miscibility aspects that can occur for two-component mixtures, our main focus was on the atomic erbium-dysprosium 
($^{168}$Er-$^{164}$Dy) and dysprosium-dysprosium ($^{164}$Dy-$^{162}$Dy) mixtures.  
Our analysis for pure-dipolar BEC was limited to coupled systems confined in pancake-type traps, after considering 
a study on the stability regime of such systems. In case of non-dipolar systems with repulsive contact intneeractions
we are able to extend the miscibility analysis to coupled systems with cigar-type symmetries.
For a coupled condensate with repulsive inter- and intra-species two-body interactions, confined by an
external harmonic trap, the transition from a miscible to an immiscible phase is verified to be much softer 
than in the case the system is confined by a symmetric hard-wall potential.
Our results, presented by density plots, are pointing out the main role of the trap symmetry and inter-species interaction 
for the miscibility.
A relevant parameter to measure the overlap between the two densities was defined and found appropriate to 
quantify the miscibility of a coupled system.
\end{abstract}
\maketitle
\section{Introduction}
\label{secI}
Following the first experimental realization of Bose-Einstein condensation (BEC) in 1995~\cite{1995-bec-exp}, 
substantial progress has been verified in experimental and theoretical investigations with ultracold quantum gases,
which can be traced by several paper reviews on the subject (as references in \cite{2002-08-becrev}), as 
well as by a recent book written by experts in the field~\cite{2017-proukakis}. One can also follow the advances in 
quantum simulations and control of BECs in reviews such as \cite{2012-block,2010-chin}.
From studies with ultra-cold atoms, we can improve our understanding on quantum properties of a large variety 
of bosonic and fermionic systems, as well as molecular configurations with different atomic species. By controlling 
atomic properties potential technological applications exist from ultra-precise clocks till quantum computation. 
In this respect, by considering BEC mixing with different atomic and molecular configurations, also with degenerate 
complex atoms (alkaline-earth, lanthanides) and Fermi gases, one can investigate the crossover of BEC properties 
with Bardeen-Cooper-Schrieffer (BCS) superconductivity, superfluid to Mott insulator transition in bosonic and fermionic 
systems, quantum phases of matter in optical lattices, ground-state fermionic molecules~\cite{bec-works}. 

BEC with two-components was first produced with different hyperfine states of $^{87}$Rb~\cite{Myatt1997}. This is a 
simple example of a multicomponent system made with a mixture of two-species of bosons. Following that, one should 
notice several investigations using binary mixtures. This can be exemplified with works considering the dynamics of 
binary mixtures with bosons and fermions~\cite{2004-chui}; studies on the dynamics of phase separation and on how 
to control it~\cite{1998-Hall-Phase-sep,2010-Tojo-phasecontrol}; also considering mixtures with different isotopes of the 
same atomic species~\cite{2008-Papp-Tune-Mis}, or with different atomic species~\cite{K-Rb-Cs}. 
As a relevant characteristics of multicomponent ultracold gases, we have their miscibility behavior which will depend 
on the nature of the interatomic interactions between different species.
Miscible or immiscible two-component BEC systems can be distinguished by the spatial overlap or separation of 
the respective wave-functions of each component. Their  phase separations were observed in spinor BECs of sodium 
in all hyperfine states of $F=1$~\cite{hyperfine-sodium}. The advances in the experimental investigations with 
multi-component BECs have activated a large amount of theoretical descriptions applied to condensed mixtures 
having spatially segregated phases, by studying their properties related to static and dynamical 
stability~\cite{Law-1997,Chui-1998,2002-Barankov-mix,2008-Abdu,2011-Pasquiou,2012-Wilson,phase-param,
pattinson,salerno-2014}.

Dipolar atomic and molecular systems, as well as mixtures with different dipolar atoms, have been explored in 
theoretical works connected with BEC since 2000~\cite{Yi2000,Santos2000}, followed by several other investigations 
motivated by the increasing experimental possibilities in cold-atom laboratories to probe the theoretical analysis and 
eventual proposals. The theoretical effort in this direction can be traced back by the following sample works,
Refs.~\cite{2002-Goral,2002-Giovanazzi,2005-Malomed,2006-Bortolotti,2007-Ronen,2008-Koch,2008-Lahaye,
2008-Wilson,Malomed-2010,Zaman2011,martin-2012,Adhikari-2012,2013-Bisset}. A more complete bibliography 
on the subject, covering experimental and theoretical approaches, can be found in recent reviews and dissertations, 
as given in Refs.~\cite{2008-Baranov,2009-Lahaye,2010-Ueda,2013-Bisset,2012-Aika-er}. 

When considering the investigations with dipolar systems in cold-atom laboratories, 
the pioneer work is the experiment with chromium $^{52}$Cr, reported in Ref.~\cite{2005-Gries-cr}.
Among the works following that, we have the investigations on the stability of dipolar gases~\cite{2008-Koch}, 
as well as on collisions of dipolar molecules~\cite{2010-Ni}.
With ultra-cold atoms having non-negligible magnetic moments, we have experiments with dysprosium 
described in Ref.~\cite{2010-Lu-Dy}, considering the dysprosium isotopes $^{162,164}$Dy; 
and the investigation with erbium $^{168}$Er in Ref.~\cite{2012-Aika-er} .  More recently, quantum droplets have 
being observed in a strongly dipolar condensed gas of $^{164}$Dy~\cite{2016-Ferrier}, with new features being 
verified for dipolar BECs, due to the competition between isotropic short-range contact interaction and anisotropic 
long-range dipole-dipole interaction (DDI).

By considering the quite interesting recent investigations in ultracold laboratories with
two-component dipolar BECs, studies on stability and miscibility properties are of interest due to the number of 
control parameters that can be explored in new experimental setups. 
The parameters are given by the strength of dipoles, number of atoms in each component, 
inter- and intra-species scattering lengths, as well as confining trap geometries or optical lattices. 
Among the theoretical studies cited above (most concentrated on stability of dipolar condensates), we have 
some of them are particularly related to miscibility of coupled BECs and structure formation, as  
Refs.~\cite{2005-Malomed,Malomed-2010,2012-Wilson,Adhikari-2012}.

In the present paper, our main proposal is to discuss miscibility conditions for general three-dimensional (3D) 
atomic BEC systems, which are constituted by two-coupled  dipolar or non-dipolar species confined by asymmetric 
cylindrical harmonic traps. Due to stability requirements, the dipolar systems that we are considering will be confined 
in pancake-type traps. In case of non-dipolar coupled systems, we also discuss the miscibility by considering cigar-type symmetries.
For our study on the miscibility, we start with a brief discussion by considering the homogeneous case. 
In order to simplify the formalism and a possible experimental realization, both species are assumed to be confined 
by a cylindrical trap with the same aspect ratio. Without losing the general conclusions related to miscibility of 
two-atomic dipolar BEC species, most of our study will focus on the particular $^{168}$Er-$^{164}$Dy and 
$^{164}$Dy-$^{162}$Dy mixtures, motivated by the actual experimental possibilities~\cite{2012-Aika-er,2010-Lu-Dy}.

Our numerical results for the coupled dipolar Gross-Pitaevskii (GP) equation are presented by using different parameter 
configurations for the trapping properties, as well as for the inter- and intra-species two-body contact and dipolar interactions. 
The parameter region of stability for the dipolar system is discussed for different trap-aspect ratio and 
 number of atoms in each species. As we are going to evidence, for a given mixture of two condensates confined by 
 harmonic traps, the main parameters of the system that are possible to be manageable in an experimental realization 
 with focus on the miscibility are the trap-aspect ratio and the two-body scattering lengths (these ones, controlled via 
 Feshbach resonance techniques~\cite{1998-inouye,2010-chin}).

Within our full-3D model for the coupled densities, when considering pure dipolar systems trapped in pancake-shaped 
harmonic potentials, we are also discussing some unusual local minimum structures and fluctuations in the densities, 
which are verified when the system is near the instability border (considering the critical aspect ratio and atom numbers).  
These structures are verified for coupled systems that are partially immiscible ($^{168}$Er-$^{164}$Dy, in the present case), 
as well as when it is completely miscible, such as $^{164}$Dy-$^{162}$Dy. Such structures, verified for well defined trap-aspect 
ratio and number of atoms in stable configurations, suggest possible experimental studies with two-component dipolar BECs, 
considering miscible and immiscible systems.

The next sections are organized as follows. In Sec.~\ref{secII}, we present the general 3D mean-field formalism
 (in full-dimension and  dimensionless) for trapped two-component dipolar BECs, together with the definition of 
relevant parameters, as well as the numerical approach we are considering. In Sec.~\ref{secIII}, we first write down the miscibility 
conditions for homogeneous coupled systems, followed by the definition of an appropriate miscibility parameter, which is
found appropriate to measure the overlap between densities of a general coupled system. 
Our numerical results are organized in two sections, in order to characterize the main relevant conditions for
the observation of miscibility in coupled BEC systems. 
The role of the trap symmetry for the miscibility, considering different dipolar and non-dipolar mixed systems, is
analyzed in Sec.~\ref{secIV}.
In view of the particular relevance of the inter-species two-body interactions on the miscibility, the corresponding 
results are presented and discussed in Sec.~\ref{secV}. 
Finally, in Sec.~\ref{secVI}, we present a summary with our principal conclusions and perspectives.

\section{Formalism for coupled BEC with dipolar interactions}
\label{secII}
For a mixed system with two atomic species identified by $i=1,2$, having their masses, number of 
particles and local time-dependent wave-functions given by $m_i$, $N_i$, and $\psi_i\equiv  \psi_i({\mathbf r},t)$,
respectively, the general form of the mean-field GP equation, for the trapped system with dipolar interactions,  
can be described by~\cite{2002-Goral},
{\small
\begin{align}
{\rm i} \hbar \frac{\partial \psi_i({\mathbf r},t)}{\partial t} 
&= \,\bigg [ -\frac{\hbar^2}{2m_i}\nabla^2+V_i({\mathbf r})+ \sum_{j=1}^2 G_{ij}N_j \vert \psi_j({\mathbf r},t) \vert^2  
\label{eq_DBEC1}\\ 
& \, +  \sum_{j=1}^2 \frac{N_j}{4\pi} 
\int d^3{\mathbf r'} V_{ij}^{(d)}({\mathbf r - \mathbf r'}) \vert\psi_j({\mathbf r'},t)\vert^2  
{\bigg ]}  \psi_i({\mathbf r},t)\notag
\end{align}
}
where $V_i(\mathbf r)$ is the trap potential for each species $i$, with $V_{ij}^{(d)}(\mathbf{r- r'})$ defining 
the magnetic-type dipolar interaction between particles $i$ and $j$.
The nonlinear contact interactions between the particles are given by 
$G_{ij}\equiv({2\pi \hbar^2}/{m_{ij}})a_{ij}$, where $a_{11}$, $a_{22}$ and $a_{12}=a_{21}$
are the two-body scattering lengths for intra ($a_{ii}$) and inter ($a_{12}$) species, 
with $m_{ij}$ being the reduced mass $m_im_j/(m_i+m_j)$. 
In the above, both wave-function components are normalized as
\begin{equation}
\int d^3{\mathbf r} \vert \psi_{i}({\mathbf r},t) \vert^2 = 1.
\label{norm}\end{equation}
For the confining trap potentials we assume harmonic cylindrical shapes, with frequencies $\omega_i$ 
and aspect ratios $\lambda_i$, such that 
\begin{equation}
V_i({\mathbf r})=\frac{m_i \omega_i^2}{2} (r_1^2+r_2^2+\lambda^2_i r_3^2 ),
\label{trap}\end{equation} 
where it will be assumed that each species $i$ is confined by an angular frequency $\omega_i$ along the 
$x-y$ plane, $\vec{\rho}\equiv r_1\hat{e}_1+ r_2 \hat{e}_2$; and with $\lambda_i \omega_i$ 
along the $z-$direction $r_3\hat{e}_3$.  The trap will be spherically symmetric for $\lambda_i=1$; will have 
a cigar shape for $\lambda_i<1$; and a pancake shape when $\lambda_i >1$.

For the magnetic-type dipolar interaction between particles $i$ and $j$, with respective dipole momentum 
strength given by $D_{ij}\equiv\mu_0\mu_i\mu_j$ ($\mu_0$ being the permeability in free space and $\mu_i$ 
the dipole moment of the species $i$), we have 
\begin{align}
V_{ij}^{(d)}({\bf r-r'}) = D_{ij} \frac{1-3\cos^2 \theta}{ \vert  {\bf r-r'} \vert  ^3} 
\label{Vdd}\end{align}
where ${\bf r -r'}$ determines the relative position of dipoles and $\theta$ is the angle between ${\bf r-r'}$ and 
the direction of polarization.  

Let us rewrite (\ref{eq_DBEC1}) in dimensionless quantities, with the first component defining the scales 
for length, with  $l\equiv\sqrt{{\hbar}/{(m_1\omega_1 )}}$ and energy, $\hbar\omega_1$. Within these units, we 
introduce new dimensionless variables and redefine the parameters such that 
\begin{eqnarray}
{\mathbf x}&\equiv&\frac{\mathbf r}{l} = x\hat{e}_1+y\hat{e}_2+z\hat{e}_3\equiv \vec{\rho}+z\hat{e}_3,\nonumber\\
\nabla_{\mathbf x}&=&l^3\nabla_{\mathbf r},\;\; \tau\equiv\omega_1 t,\nonumber\\
g_{ij}&\equiv&\frac{G_{ij}N_j }{\hbar\omega_1 l^3}=\frac{2\pi m_1 a_{ij} N_j}{m_{ij} l},\;\; 
\sigma\equiv\frac{ m_2\omega_2^2}{m_1\omega_1^2},\\ 
a^{(d)}_{ii}&=&\frac{D_{ii}}{12\pi }\frac{m_i}{m_1}\frac{1}{\hbar\omega_1l^2},\;
a^{(d)}_{12}=a^{(d)}_{21}= \frac{D_{12}}{12\pi}\frac{1}{\hbar\omega_1l^2},\nonumber \\
d_{ij}&=& \frac{N_j D_{ij}}{4\pi}\frac{1}{\hbar\omega_1\,l^3}
.\nonumber
\end{eqnarray}
With the above,  and also by redefining the wave-functions for the atomic species $i$, $\phi_i({\mathbf x},\tau) = 
\sqrt{l^3}\;\psi_i({\mathbf r},t)$, the expression (\ref{eq_DBEC1}) can be rewritten as the following dimensionless 
coupled expressions: 
{\small 
\begin{align}
{\rm i} \frac{\partial \phi_1}{\partial\tau}=& \,{\bigg [}-\frac{1}{2}{\nabla_{\bf x}^2}+ \frac{1 }{2} (\rho^2+\lambda^2_1 z^2 ) 
+ g_{11} \vert \phi_1 \vert^2   
+ g_{12} \vert \phi_2 \vert^2  \notag \\ & \,
+  \int d^3{\mathbf y} \frac{1-3\cos^2 \theta}{ \vert  {\bf x-y} \vert  ^3} 
\left(d_{11}\vert\phi_1'\vert^2 +  d_{12}\vert\phi_2'\vert^2 \right) {\bigg ]}  \phi_1,
\label{eq_num_DBEC1}
\end{align}
and
\begin{align}
{\rm i} \frac{\partial \phi_2}{\partial\tau}=& \,{\bigg [} - \frac{m_{1}}{2m_2}\nabla_{\bf x}^2+\frac{\sigma}{2}(\rho^2+\lambda^2_2 z^2 ) 
+ g_{21} \vert \phi_1 \vert^2 + g_{22} \vert \phi_2 \vert^2 \notag  \\ &  
+ \int d^3{\mathbf y} \frac{1-3\cos^2 \theta}{ \vert  {\bf x-y} \vert  ^3} 
\left( d_{21}\vert\phi_1'\vert^2  + d_{22} \vert\phi_2'\vert^2 \right) 
{\bigg ]}  \phi_2
\label{eq_num_DBEC2}
,\end{align}
where $\phi_i\equiv \phi_i({\mathbf x},\tau)$ and $\phi_i'\equiv \phi_i({\mathbf y},\tau)$.
}

\subsection{Dipolar and contact interaction parameters}
In our analysis, two kind of coupled atomic system are being treated. First, with the erbium $^{168}$Er 
and dysprosium $^{164}$Dy, assumed with moment dipoles $\mu=7\mu_B$ and  $\mu=10\mu_B$, 
respectively ($\mu_B$ is the Bohr magneton).
Next, both components of the coupled system are from isotopes of the same atomic species, 
$^{164}$Dy and $^{162}$Dy, such that the moment dipoles are the same $\mu=10\mu_B$ for 
both components. As a rule we define as component 1 in the mixture the more massive atomic
species.

For the angular frequencies of the axial traps, we use $\omega_1=2\pi\times$60 s$^{-1}$ for the 
$^{168}$Er and $\omega_2=2\pi\times$61 s$^{-1}$ for $^{164}$Dy and $^{162}$Dy, 
corresponding to $\sigma\approx 1$, 
with equal aspect ratios for both components:
$\lambda\equiv\lambda_1=\lambda_2$. 
The time and space units will be such that $\omega_1^{-1}=2.65$ms and 
$l=1\mu$m ($= 10^{4}{\textup{\AA}}$  $= 1.89\times10^{4} a_0$). 
In case of purely dipolar BECs, we take all the two-body scattering lengths $a_{ij}=0$. In several other cases
we fix the scattering lengths between $10a_0$ and $110 a_0$, where $a_0$ is the Bohr radius.
In order to compare the dipolar and contact interactions, the parameters for the 
intra- and inter-species dipolar interactions are given in terms of the length scale. For the moment dipole of the species,
in terms of the Bohr magneton $\mu_B$, we assume  $\mu=7\mu_B$ for $^{168}$Er, with $\mu=10\mu_B$ for both 
species of dysprosium,  $^{164,162}$Dy.
The two-body scattering lengths $a_{ij}$ and dipolar interactions $a^{(d)}_{ij}$, given in units of $a_0$, 
are related to the corresponding dimensionless parameters $g_{ij}$ and $d_{ij}$ as: 
$a_{11}/a_0\approx1504{g_{11}}/N_1$, $a_{ij\ne11}/a_0\approx1486{g_{ij}}/N_j$, 
$a^{(d)}_{11}/a_0\approx6301{d_{11}/N_1}$, $a^{(d)}_{22}/a_0\approx 6151 {d_{22}}/N_2$, 
$a^{(d)}_{12}/a_0\approx6301 {d_{21}/N_1}$ 
$ =6301 {d_{12}/N_2}$.

\subsection{Numerical method and approach}
For the numerical approach, used to obtain our results when solving the 
full-3D coupled Eqs.~(\ref{eq_num_DBEC1}) and (\ref{eq_num_DBEC2}),
we have employed the split-step Crank-Nicolson method, which is detailed in similar non-linear studies, as 
in Refs.~\cite{Gammal2006,2008-Abdu,CPC1},  
where one can find more extended analysis and details on computer techniques convenient for nonlinear coupled 
equations, facing stability and accuracy of the results.
In view of the particular integro-differential structure of the coupled nonlinear differential equations
when having dipolar interactions, we had to combine our approach in solving coupled differential 
equations with a standard method for evaluating dipolar integrals in momentum space~\cite{2002-Goral,CPC2,CUDA}.  

By looking for stable solutions, the 3D numerical simulations were carried out in imaginary time with a 
grid size having $128$ points for each dimension, where we have $\Delta x = \Delta y = \Delta z = 0.2$ for the
space-steps and $\Delta t = 0.004$ for the time-step. The results were quite stable, verified by taking half of the 
mentioned grid sizes. 

As a preliminary calculation, which also help us to check the numerical code, we reproduce the stability 
diagram obtained in Ref.~\cite{2002-Goral} for a spherically symmetric  trap ($\lambda = 1$),
where only one atomic species was used, $^{52}$Cr, for the coupled system, with the two species ($m_1=m_2$) 
having opposite polarizations along  the $z$ direction. Besides that, in our analysis, we have also verified the stability
of the numerical results by studying the effect of varying $\lambda$. In this regard, once
verified that pancake-type configurations are required for stable dipolar configurations, the miscibility in cigar-type 
traps is being analyzed only for non-dipolar systems.

\section{Miscibility of coupled systems}
\label{secIII}
\subsection{Homogeneous case with hard-wall barriers}
In order to characterize the transition between miscible and immiscible states, let us consider a simpler case, 
for the homogeneous 3D system with $V_1({\mathbf r})=V_2({\mathbf r})=0$ and hard-wall 
barriers, following the simplified energetic approach presented in Ref.~\cite{Chui-1998}. 
Within this approximation, 
the miscible-immiscible transition (MIT) can be characterized by a threshold parameter, which is defined by the 
relation between the two-body repulsive interactions.
The criterium for miscibility, also quoted in some recent works on binary BEC mixtures (see, for instance, 
Refs.~\cite{2012-Wilson,pattinson}), was previously obtained from stability analysis of the excitation spectrum in 
Ref.\cite{Law-1997}. It can be easily generalized to include all the two-body repulsive interactions.

As in the present case we are interested in mixed configuration with contact and dipolar interactions, we should
first obtain a simple relation for the dipolar interactions appearing in the formalism. 
For that, we follow the approach given in Ref.~\cite{2002-Goral} to deal with the integro-differential formalism where 
we have a divergence of the integrand in the zero limit for the inter-particle distances. The convolution 
theorem and Fourier transforms are applied for the magnetic dipolar potential and for the density components $|\psi_i|^2$. 
The Fourier transform of the dipolar potential (\ref{Vdd}) is given in terms of a cut-off parameter, which is of the 
order of the atomic radius. As this parameter is much smaller than a significant length scale of 
the system, one can safely consider the limit where it is zero, such that  the dipolar potential between the atomic 
species $i$ and $j$ is given by
\begin{equation} 
{\cal V}_{ij}^{(d)}({\mathbf k}) = D_{ij}\frac{4\pi}{3}(3 \cos^2\theta_{\mathbf k} -1)\equiv
D_{ij}\frac{4\pi}{3}f(\theta_{\mathbf k}),
\label{ffVdd}
\end{equation}
where $\theta_{\mathbf k}$ is the angle between the wave-vector ${\mathbf k}$ and the dipole moment. 

Therefore, by including together the contact and dipolar interactions, the condition for MIT can be written as 
\begin{small}
\begin{align}
\hspace*{-.7cm}
\Delta = \frac{\left|G_{11}+D_{11}f(\theta_{\mathbf k})\right|
\left|G_{22}+D_{22}f(\theta_{\mathbf k})\right|}{\left[G_{12}+D_{12}f(\theta_{\mathbf k})\right]^2} -1
,\label{delta}
\end{align}
\end{small}
where $\Delta=0$ defines the critical value for the transition from miscible ($\Delta>0$) to immiscible 
($\Delta<0$) systems.
The mixed coupled state will have a lower total energy when the mutual repulsion between atoms is 
large enough  such that we have $\Delta < 0$, which is characterizing the system in an immiscible phase. The
system will be in a miscible phase when $\Delta > 0$, with a critical border for the transition given by 
$\Delta = 0$. The MIT occurs when the inter-species and intra-species interactions are balanced. As verified, 
the two-body interactions are just being scaled by the dipolar interactions, in this simple case that the system is 
confined by hard-wall barriers. The observation that the dipolar interactions are not playing a significant role 
results from the fact that the dipole-dipole interactions averaged out for homogeneous gases. 
However, for numerical comparisons with more general miscibility conditions, 
it is still quite useful, as it does not depend on the number of atoms and condensate size. 

\subsection{Miscibility in general coupled systems}
The miscible to immiscible transition for a coupled system, as defined by the critical limit $\Delta$ in Eq.~(\ref{delta}), 
is changed from a first-order transition that occurs for homogeneous system to a second-order one when the
kinetic energy is taken into account, as discussed in Ref.~\cite{phase-param}.  
Therefore, in order to evaluate the miscibility of a coupled system for a more general case, we define a parameter 
which gives an approximate measure of how much overlapping we have between the densities of both components.
By the definition of a parameter related to the densities we are modifying the one suggested in Ref.~\cite{phase-param}, 
where the overlap between the wave-functions was considered. Both definitions are equivalent in particular cases. 
Our parameter to define the miscibility of a coupled system is given by
\begin{align}
\eta = \int \vert \phi_1 \vert \vert \phi_2 \vert \ d{\mathbf x} = \int \sqrt{\vert \phi_1 \vert^2 \vert \phi_2 \vert^2} \ d{\mathbf x}
.\label{eta}\end{align}
As  $\phi_1$ and $\phi_2$ are both normalized to one, also having the same center, this expression implies that  
$\eta=1$ for the complete overlap between the two densities, decreasing as the overlap diminishes. 
Therefore, we can define the system as almost completely immiscible when $\eta \ll 1$ (close to zero); 
and, almost completely miscible when $\eta$ is close to one.  This parameter is extending to general non-homogeneous 
mixtures the MIT criterium (\ref{delta}) discussed for homogeneous systems.
With $\eta$, intermediate cases can be determined when the system is partially immiscible or partially miscible. 
From our observation, which will follow from the analysis of results obtained for the densities in the
next sections, when $\eta \lesssim 0.5$ the system shows already a clear space separation, with the 
components having their maxima in well separated points in the space, such that we can already define the 
coupled system as immiscible. For $\eta$ between $0.5$ and $0.8$, the two densities start having an increasing 
overlap with their maxima approaching each other, such that we can define the system within this interval as partially 
miscible. The maxima of the densities are close together for $\eta \gtrsim 0.8$, when we assume the system 
is miscible. 

As a side remark, we should observe that we are treating two condensates with different atomic species symmetrically 
distributed around the center. As a general expected behavior, the density of the more massive species should be closer 
to the center, with the other density being pushed out. 
This can be explained even before activating particular interactions between the atoms, 
as the kinetic energy of the massive species is smaller than the corresponding kinetic energy of the less massive species.

Next two sections are dedicated to present our main results on the miscibility of coupled dipolar and non-dipolar BECs. 
When choosing non-zero dipolar parameters, the results are exemplified by two mixtures for a better characterization of 
the miscibility properties. Within the actual experimental possibilities in BEC laboratories, we consider the  
erbium-dysprosium ($^{168}$Er-$^{164}$Dy) and dysprosium-dysprosium ($^{164}$Dy-$^{162}$Dy) mixtures.

\section{Miscibility results - role of the trap symmetry}
\label{secIV}
\label{sec:numerical}
The results analyzed in this section for the miscibility properties of coupled atomic condensates are more concerned
with the role of the symmetry of the harmonic trap, considering different configurations for the possible internal 
interactions (contact and dipolar) between the atoms of both species. 
The contact interactions are characterized by the atomic two-body scattering lengths,
with dipolar interactions due to the magnetic dipole moment of each atomic species. 
In the following, we split the section in three subsections for clarity. In part A, we consider miscibility in the case we 
have no contact interactions, starting with a discussion on the stability of such dipolar systems in terms of the 
trap-aspect ratio and number of atoms. As shown, for realistic number of atoms, we need pancake-shaped traps.
Next, in part B, our analysis is concentrated in non-dipolar coupled structures, where we can examine cigar-type
and pancake-type BEC configurations. We conclude the section with part C, by analyzing our results for the case 
that we have both dipolar and contact interactions. 

\subsection{Miscibility in pure dipolar interactions}
\subsubsection{Stability analysis}
%fig1
\begin{figure}[t]
\centering\includegraphics[width=0.85\columnwidth]{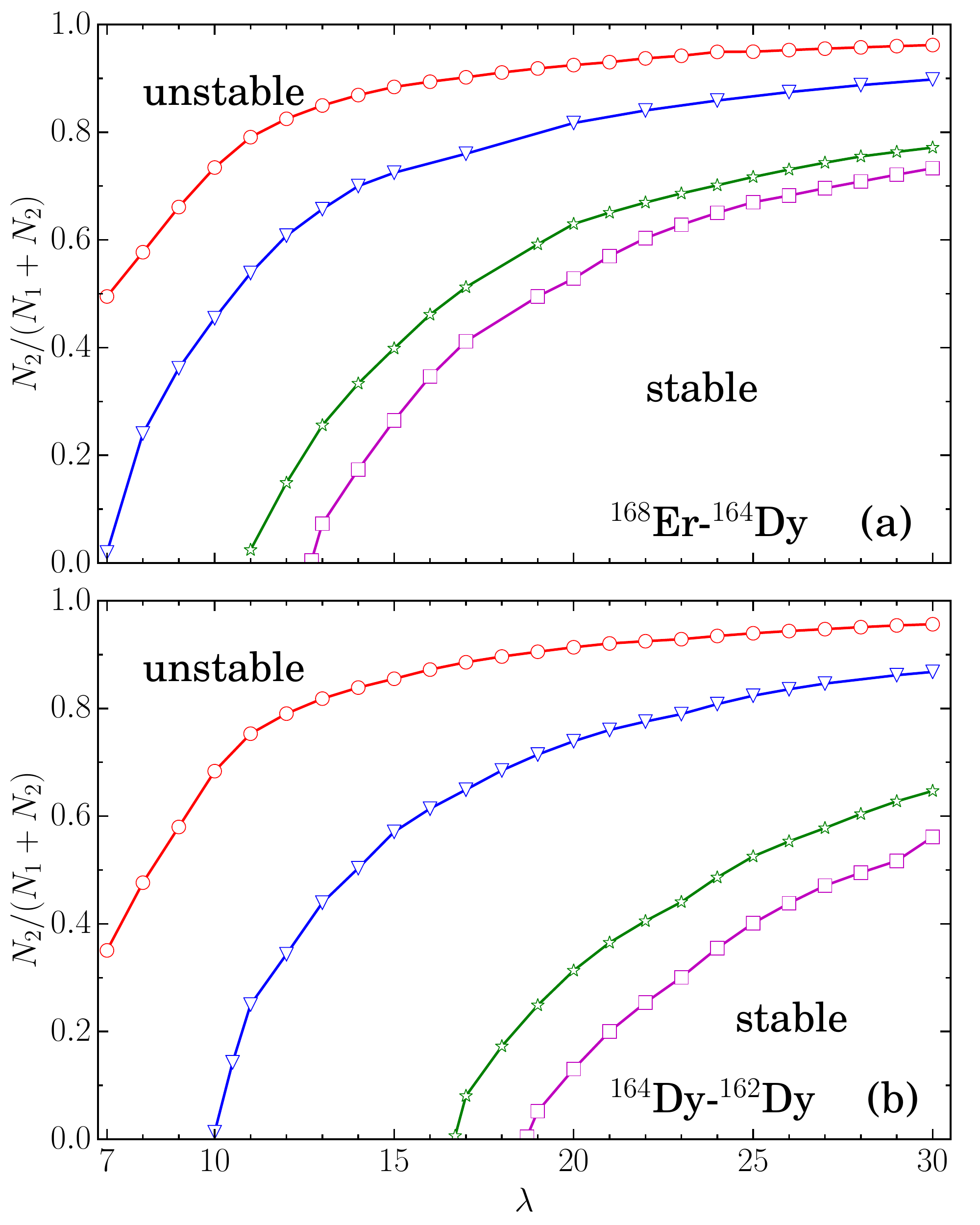}
\caption{(Color online) {Stability limits for the coupled pure-dipolar ($a_{ij} = 0$) BEC systems, 
$^{168}$Er-$^{164}$Dy (upper panel, where $^{168}$Er is the species 1) and $^{164}$Dy-$^{162}$Dy (lower 
panel, where $^{164}$Dy is the species 1), showing the critical fraction of atoms $N_2/N$ as functions of the trap 
aspect ratio ($\lambda$).
In both panels, the given lines refer to four fixed values for the number of atoms of the species 1: 
$N_1$=1000 (red line with circles), $N_1$=3000  (blue line with triangles), $N_1$=8000  (green line with stars)  and $N_1$=10000  
(magenta line with squares). For each one of the critical curves, only the lower-right region gives stable BECs.
For the dipolar parameters, we have
$a^{(d)}_{11}=66a_0$,  $a^{(d)}_{22}= 131a_0$  and $a^{(d)}_{12}=94a_0$, in panel (a); and  
$a^{(d)}_{11}=132a_0$, $a^{(d)}_{22}= 131a_0$  and $a^{(d)}_{12}=131a_0$ in panel (b).
 }
}
\label{fig1}
\end{figure}
The homogeneous case of purely dipolar condensate is unstable due to anisotropy of the 
DDI.  However, in a way similar as the case of homogeneous non-dipolar BEC that are unstable
(for attractive two-body interactions, $a<0$), the instability usually can be overcome by applying 
some external trap, which  will help to stabilize the dipolar BEC by imprinting anisotropy to the 
density distribution~\cite{2008-Koch}. This is applicable for single and multi-component dipolar BECs. 
On the stabiliy of two condensates purely dipolar, separated by a distance, we can mention 
Ref.~\cite{Zaman2011}. For the case of single component dipolar condensates, it was previously 
shown in Refs.~\cite{Yi2000,Santos2000} that, by increasing the aspect ratio $\lambda$ one can obtain 
a more stable configuration due to the dipole-dipole interaction becoming effectively more repulsive. 

As shown by the stability diagrams given in the two panels of Fig.~\ref{fig1}, pure dipolar condensates 
require pancake-type traps to be stable, with the coupled mixture becoming less stable when the 
dipolar strengths (inter and intra-species) are close to the same values (for some fixed ratio of number of
atoms in both species). This effect is mainly due to the inter-species repulsion in comparison with the 
corresponding intra-species ones.

To become clear this effect, our results are given for the coupled equations~(\ref{eq_num_DBEC1}) 
and (\ref{eq_num_DBEC2}), considering the dipolar BEC mixtures with $^{168}$Er-$^{164}$Dy  (upper panel) , where $^{168}$Er (1st component) and $^{164}$Dy 
(2nd component) and $^{164}$Dy-$^{162}$Dy (lower panel), where $^{164}$Dy (1st component) and 
$^{162}$Dy (2nd component).
The fraction number of atoms $N_2/N$, where $N\equiv(N_1+N_2)$ is shown as a function of 
the aspect ratio $\lambda$, 
considering four sample fixed values for the component 1 of the mixture, which are given by 
$N_1=$1000 (red lines with circles), 3000 (blue lines with triangles), 8000 (green lines with stars)
and 10000 (magenta lines with squares). In both the cases, we use purely dipolar BECs ($a_{ij}=0$). 
The dipolar parameters of the $^{168}$Er-$^{164}$Dy coupled system are
$a^{(d)}_{11}=66a_0$, $a^{(d)}_{22}= 131a_0$  and $a^{(d)}_{12}=94a_0$. Also, the dipolar parameters of 
the $^{164}$Dy-$^{162}$Dy mixtures are 
$a^{(d)}_{11}=132a_0$, $a^{(d)}_{22}= 131a_0$  and $a^{(d)}_{12}=131a_0$.
From both the panels, one can extract the information that the stability of pure-dipolar mixtures is mainly 
affected by the inter-species strengths of the dipolar interactions (in comparison with the corresponding 
intra-species strengths). The systems are more stable if less repulsion occurs between inter-species atoms. 
By comparing the lower with the upper panel, we can
verify the effect of reducing by about half the dipolar strength of one of the component, increasing
the stability of the system. The maximum effect can be seen for $N_2=0$ ($N=N_1$), implying that
a system with 10000 atoms of $^{168}$Er can only be stable within a pancake-like trap with
$\lambda \gtrsim 13$, whereas with the same number of $^{164}$Dy atoms the stability 
can only be reached for $\lambda\gtrsim 19$.

For a fixed aspect ratio $\lambda$,  the two-component BEC can become 
unstable by increasing the fraction $N_2/N$, where the critical number varies according to the fraction $N_1/N$. 
Also, there is a critical trap aspect ratio ($\lambda_c$) for the stability,  
As one can verify from the upper panel, for the $^{168}$Er-$^{164}$Dy mixture with  
 $N_1$\,= 3000, 8000, and 10000 the, this critical aspect ratio starts from  
$\lambda_c\, \approx\,$ 7, 11, and 13  respectively. On the other case, for the $^{164}$Dy-$^{162}$Dy mixture 
with the same sets of $N_1$( \,= 3000, 8000, and 10000), the critical lower limit for stability starts with 
$\lambda_c\, \approx\,$ 10, 17, and 19  respectively. This variation in the $\lambda_c$  is obviously explained 
by the difference in the dipolar strengths of both the cases, with the  $^{168}$Er component having about half of 
the dipolar strength of $^{162,164}$Dy.

From the upper panel in Fig.~\ref{fig1}, for the same total number of atoms of $^{168}$Er-$^{164}$Dy mixture, 
one can verify that, for stability, $\lambda_c$ is reduced (less deformed pancake-type trap) when considering 
$N_1>N_2$; implying larger fraction of $^{168}$Er atoms. As an example, for $(N_1,N_2)=(3000,8000)$ we have 
$\lambda_c\approx16$; and for $(N_1,N_2)=(8000,3000)$ we have $\lambda_c \approx 14$.
This behavior results from the respective strengths of the dipolar interactions of both components, 
with the erbium component having $a^{(d)}_{11}=66a_0$, which is smaller than the corresponding 
value for the dysprosium component ($a^{(d)}_{22}=131a_0$).
For the other case shown in the lower panel of Fig.~\ref{fig1}, as  both components have about equal DDI 
strengths, $\lambda_c$ depends essentially only on the total number of atoms.

This stability analysis will be considered in the following study on the miscibility of two coupled mixtures. However, 
it can also be of interest for corresponding experimental investigations (for the specific mixtures we have considered 
or for other similar dipolar mixtures).

\subsubsection{Structure of coupled pure-dipolar condensates} 
\begin{figure*}[thpb]
\begin{center}
{\LARGE \bf $^{168}$Er-$^{164}$Dy} \\
\includegraphics[width=16cm,height=9cm]{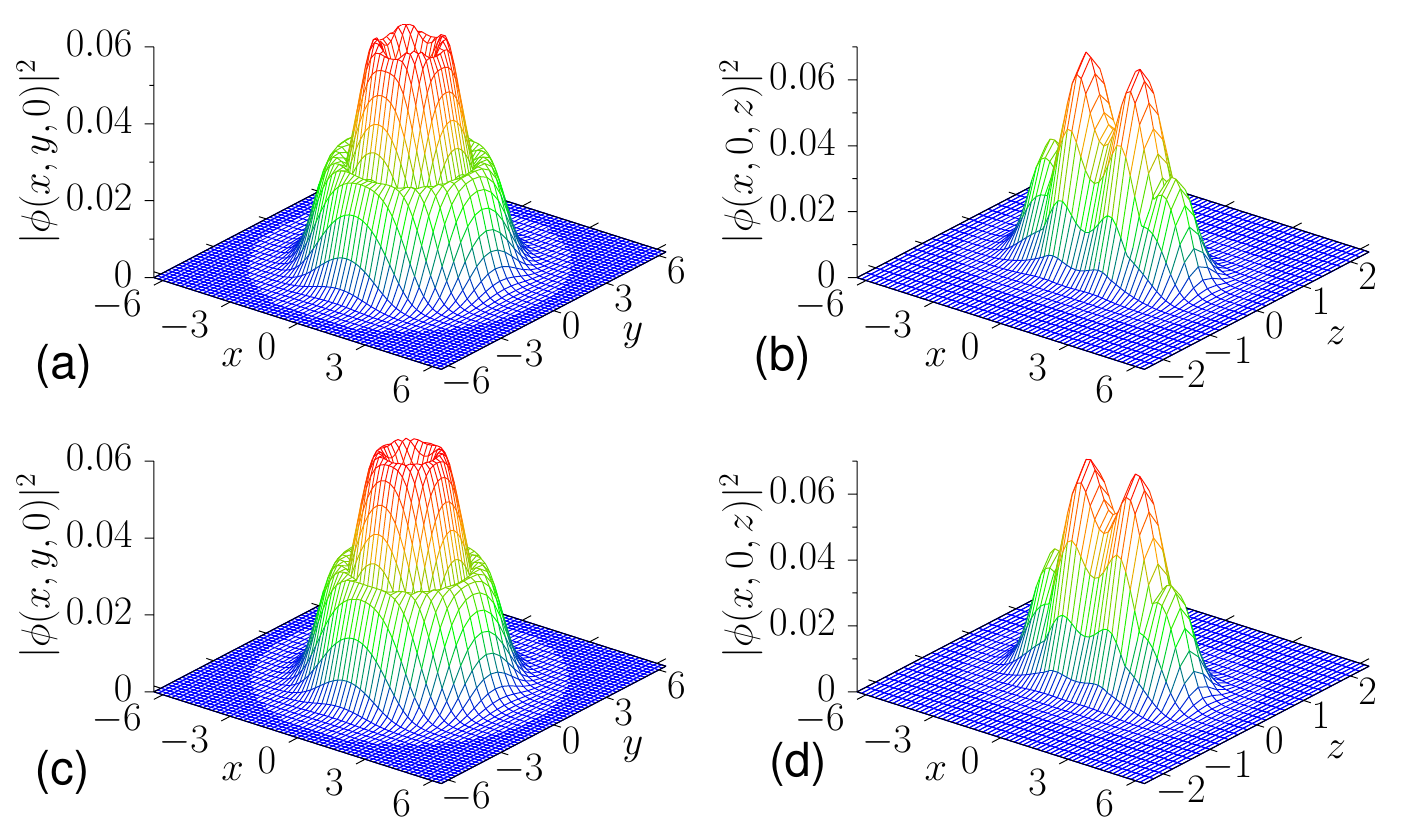}
\vskip 1cm
{\LARGE \bf $^{164}$Dy-$^{162}$Dy} \\
\includegraphics[width=16cm,height=9cm]{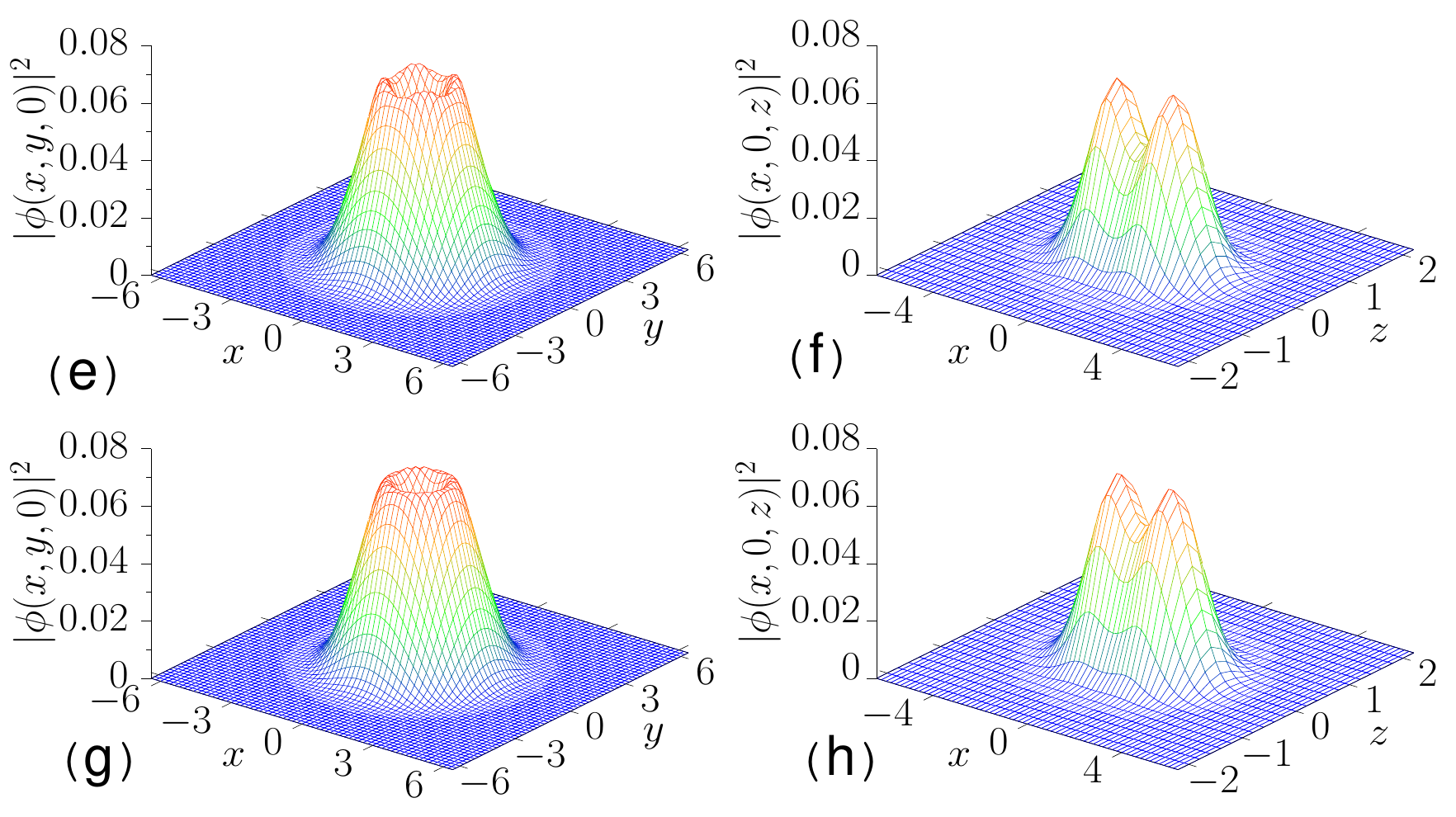}
\caption{(Color online)  Surface plots of the densities for two cases of pure-dipolar 
($a_{ij}=0$) coupled systems, confined by traps with aspect ratio $\lambda=7$. In the left panels 
we have the densities for $z=0$ in the ($x,y$) surface, with the right ones for $y=0$ in the ($x,z$) surface.  
In panels (a-d) we have the $^{168}$Er-$^{164}$Dy; considering $N_1=3000$ and $N_2=60$ in the panels (a-b)
and  $N_1=2700$ and $N_2=40$ in the panels (c-d). 
The internal (reddish) structure is dominated by the $^{168}$Er, with the surrounding (greenish) by the $^{164}$Dy. 
In panels (e-h) we have the $^{164}$Dy-$^{162}$Dy; with $N_1=1500$ and $N_2=50$ in the panels (e-f) and 
$N_1=1350$ and $N_2=50$ in the panels (g-h). 
The internal (reddish) structure is dominated by the $^{164}$Dy, with the surrounding (greenish) by the $^{162}$Dy. 
In the case of $^{168}$Er-$^{164}$Dy, shown in (a-d), the magnetic dipolar parameters are $a^{(d)}_{11}=66a_0$, 
$a^{(d)}_{22}=131a_0$ and $a^{(d)}_{12}=a^{(d)}_{21}=94a_0$. For both configurations (a-b) and (c-d) we 
obtain $\eta\approx 0.77$.
In the case of $^{164}$Dy-$^{162}$Dy, shown in (e-h), the interactions between particles are about the same, 
$a^{(d)}_{ij}=131a_0$. We obtain  $\eta\approx 0.99$ for both configurations (e-f) and (g-h). 
} 
\label{fig2}
\end{center}
\end{figure*}

Once analyzed the stability for pure dipolar coupled systems, in this subsection we characterize the 
role of the trapping aspect ratio $\lambda$ in the structure and miscibility of a pure-dipolar coupled condensate.
Therefore, as in the preceding subsection, all the two-body scattering lengths ($a_{ij}=0$) are fixed to zero and
we consider both coupled systems with $^{168}$Er-$^{164}$Dy and $^{164}$Dy-$^{162}$Dy. 
Our numerical results for the structure of the coupled system can be visualized through density plots, with  
the immiscible and miscible regimes characterized by the parameter $\eta$ defined by Eq.~(\ref{eta}). 
In Fig.~\ref{fig2} we present 3D surface plots for the densities $|\phi(x,y,0)|^2$ in the 
[frames (a), (c), (e), (g)]  and $|\phi(x,0,z)|^2$  [frames (b), (d), (f), (h)], given respectively in the 
$(x,y)$ and $(x,z)$ surfaces.
These plots are providing a 3D visualization of the density overlapping and distribution of the two-components. 
More close to the center, we have the more massive component (in red) of the mixture, which is enhanced 
when the system is more immiscible. How close to the center is the other species (in green) will depend on 
how miscible is the mixture, being clearly identified for immiscible mixtures. This can be verified by 
comparing the set of panels (a-d) for $^{168}$Er-$^{164}$Dy with the ones (e-h) for $^{164}$Dy-$^{162}$Dy.

\begin{figure*}[htpb]
\begin{center}
\includegraphics[width=0.49\textwidth]{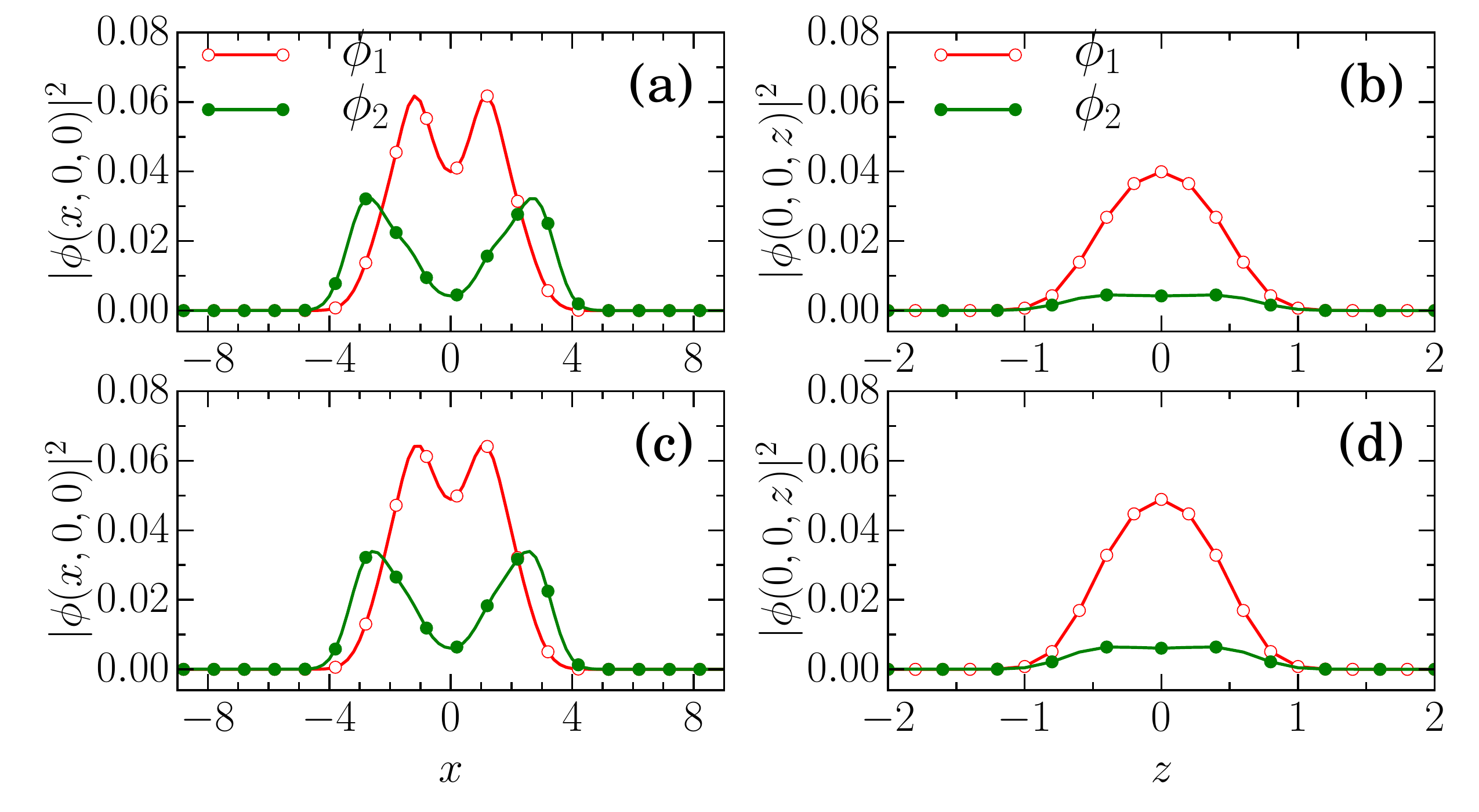}
\includegraphics[width=0.49\textwidth]{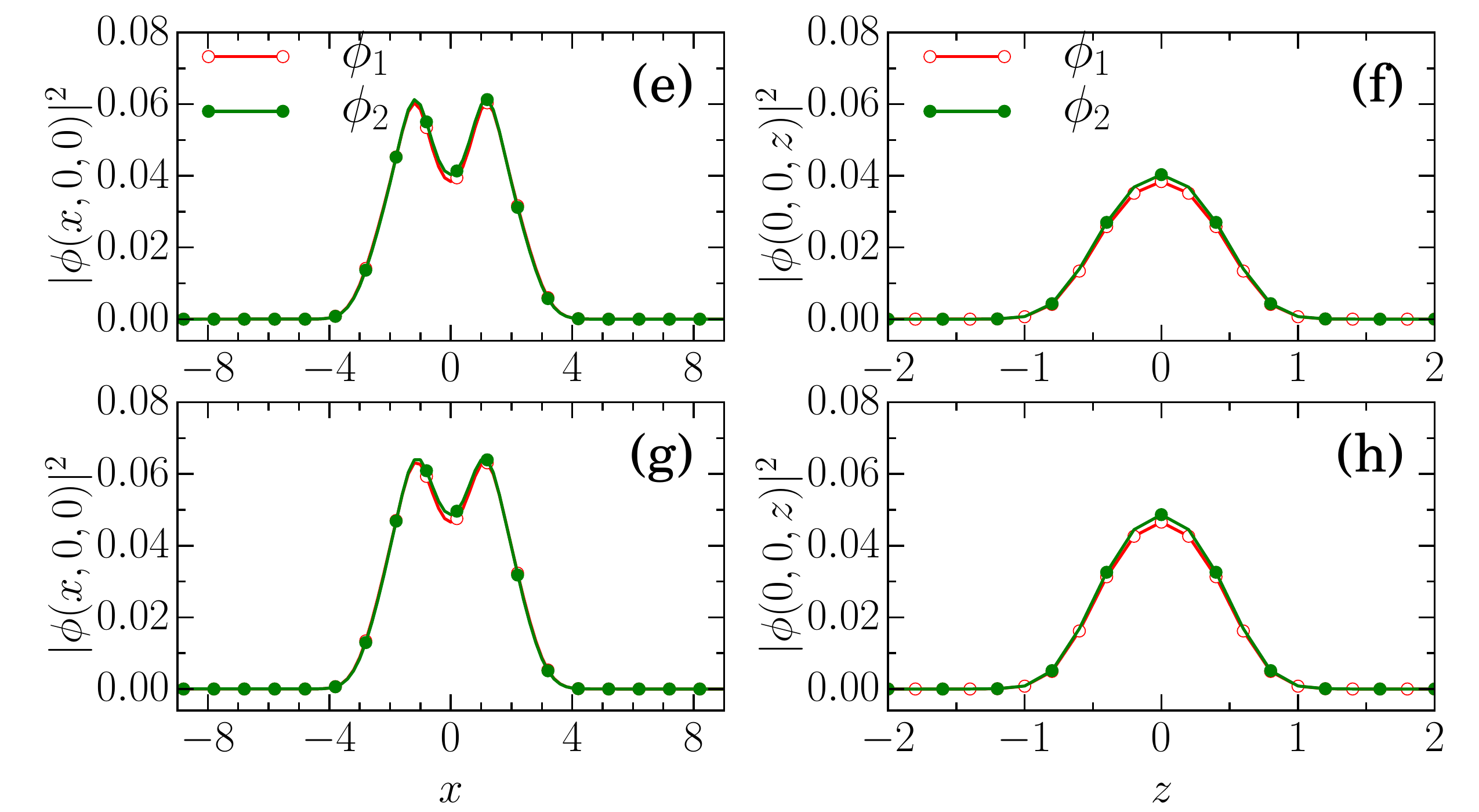}
\caption{(Color online) For the parameterizations used in Fig.~\ref{fig2}, we have the corresponding 1D plots for the 
central densities, as functions of $x$ and $z$. The panels are in direct correspondence with the ones of Fig.~\ref{fig2}.
The $^{168}$Er-$^{164}$Dy mixed system is shown in the left four (a-d) panels, with $^{164}$Dy-$^{162}$Dy given in 
the four right ones (e-h).   
The first component is shown by red lines with empty circles, with the second shown by green lines with solid circles.
} 
\label{fig3}
\end{center}
\end{figure*}

Corresponding to the panels of Fig.~\ref{fig2}, we also have the one-dimensional (1D) plots for the densities
in Fig.~\ref{fig3}, which are given as functions of one of the dimensions, $x$ or $z$, with the other two dimensions 
at the center. From these 1D densities,  the amount of overlapping between the densities can better be observed, 
being more helpful when comparing different parameter configurations.
Both systems,  $^{168}$Er-$^{164}$Dy and $^{164}$Dy-$^{162}$Dy, represented in Figs.~\ref{fig2} and \ref{fig3},
are been shown for pancake-shaped traps with $\lambda=7$. The number of atoms that was considered
was dictated by the stability of the mixtures (as one can follow from Fig.~\ref{fig1}), such that we can study 
the density properties when the mixtures are stable but not far from the border where they can become unstable.

%fig2

%fig3

In Fig.~\ref{fig2}, panels (a-d), for the $^{168}$Er-$^{164}$Dy mixture we use two different set of parameters for the 
dipolar interactions, such that we have the same value for the miscibility parameter ($\eta=0.77$). In the upper frames 
(a-b) we have $d_{11}=31.42$, $d_{12}=0.89$,  $d_{22}=1.278$ and $d_{21}=44.76$, with $N_1=3000$ and $N_2=60$; 
in the lower frames (c-d), $d_{11}=28.2$, $d_{12}=0.59$, $d_{22}=0.85$ and $d_{21}=40.2$, with $N_1=2700$ and 
$N_2=40$. 
The same densities are shown in 1D plots, in the four panels (a-d) of Fig.~\ref{fig3}, in correspondence with the 
panels (a-d) of  Fig.~\ref{fig2}, to improve the visualization of the overlap between densities.
Both components of the densities are given as functions of $x$ (left panels) and $z$ (right panels), with the other 
dimensions at the center. Note that the condensate is confined in a pancake-type cylindrical trap, with aspect ratio 
$\lambda=7$, such that the distribution along the $z-$axis is more concentrated near the center ($z=0$) than the 
other two directions. As also characterized by the miscibility parameter, $\eta=0.77$, in the present case the coupled 
system is partially miscible. This is quite well represented in the left four panels given in Fig.~\ref{fig3}.

For the coupled system $^{164}$Dy-$^{162}$Dy we use two other different set of parameters in our analysis, showing 
the densities of the two coupled components in 3D and 1D plots in the panels (e-h) of Figs.~\ref{fig2} and \ref{fig3}, 
respectively. These results are given in correspondence with the results presented for the system $^{168}$Er-$^{164}$Dy 
in the panels (a-d) of Figs.~\ref{fig2} and Fig.~\ref{fig3}.
For the parameters, we have $d_{11}=31.42$, $d_{12}=1.040$,  $d_{22}=1.065$ and $d_{21}=31.190$, with $N_1=1500$ and 
$N_2=50$, in the frames (e-f) of Figs.~\ref{fig2} and \ref{fig3};
with 
$d_{11}=28.2$, $d_{12}=1.040$, $d_{22}=1.065$ and $d_{21}=28.071$, with $N_1=1350$ and $N_2=50$,
in the frames (g-h) of these figures.
In this case, where we have two isotopes of the same atom, we notice that the system is completely miscible, having 
the same value close to one for the miscibility parameter $\eta=0.99$. The complete overlap between the densities of the two components are clearly shown in panels (e-h) of Fig.~\ref{fig3}.

In these Figs.~\ref{fig2}-\ref{fig3}, by using the aspect ratio $\lambda=7$, for each one of the mixtures we can identify a non-trivial structure 
emerging in the condensate near the boundary of stability, with a local minimum at the center (for both components) in the symmetrical 
$x-y$ plane (see left panels of both figures). At the $z-$direction, we have a normal Gaussian shape, as seen in the right panels of the figures.

In both the mixtures shown in Figs.~\ref{fig2} and \ref{fig3}, the upper panels (a-b) and (e-f) are for parameters very close to the stability threshold, 
as indicated by the results given in Fig.~\ref{fig1}. Therefore, the structure observed for the density of the component 1, which has the largest 
fraction of atoms in the coupled system, can be explained by fluctuations close to the instability regime. For each system, by going from smaller values, shown in panels (c-d) and (g-h), to larger values, shown in panels (a-b) and (e-f), of $N_1$ (keeping $N_2$ about the same), we are approaching the 
unstable regime, visualized by the occurrence of oscillation peaks in the density of the dominant component in the system. 
The observed number of four peaks around the center is related to stability requirements when the number of atoms is close to the maximum 
allowed limit for a given trap asymmetry. By small variations of the parameters near the stability, before the collapse of the
system, it is possible to increase such number of peaks.

In both sets of systems shown in Figs.~\ref{fig2} and \ref{fig3}, while observing the peak oscillations occurring for the first component, 
one can observe only the biconcave-shaped condensate for the second component. 
In this regard, we should also note that biconcave-shaped structures with local minimum in dipolar BECs have already been 
reported in the cases of single component condensates. 
In Ref.~\cite{2007-Ronen}, such structures are explained as due to roton instability for certain specific pancake-type trap 
aspect ratios $\lambda$ ($\approx$ 7, 11, 15, 19, ...,), being not observed for other values of $\lambda$. 
In our analysis, we confirm the values reported in \cite{2007-Ronen}. However, when considering single-component 
BECs, we have also verified these type of structures for other particular trap-aspect ratios, such as 
$\lambda\approx$  8, 12, 16, and 20.  However, for the cases of fully anisotropic traps, where two 
aspect ratios are considered, this kind of density fluctuation has also been observed in Ref.~\cite{martin-2012}, 
for single component dipolar BECs. 
When considering two-component systems, we have verified this kind of structured states with no particular restriction 
on the values of $\lambda$. 

The biconcave structure occurs for particular aspect ratios of the trap, due to the repulsive interaction between 
the dipoles. In the present study for a coupled dipolar system, we observe that we have also the repulsive inter-species 
DDI in addition to the intra-species DDI, leading to the observed four peaks in the density oscillations.
The biconcave structures with a local minimum and few peaks are occurring near unstable regimes, which makes the 
experimental reproduction non-trivial to be observed.

%fig4
\begin{figure*}[htpb]
\begin{center}
\includegraphics[width=0.9\textwidth]{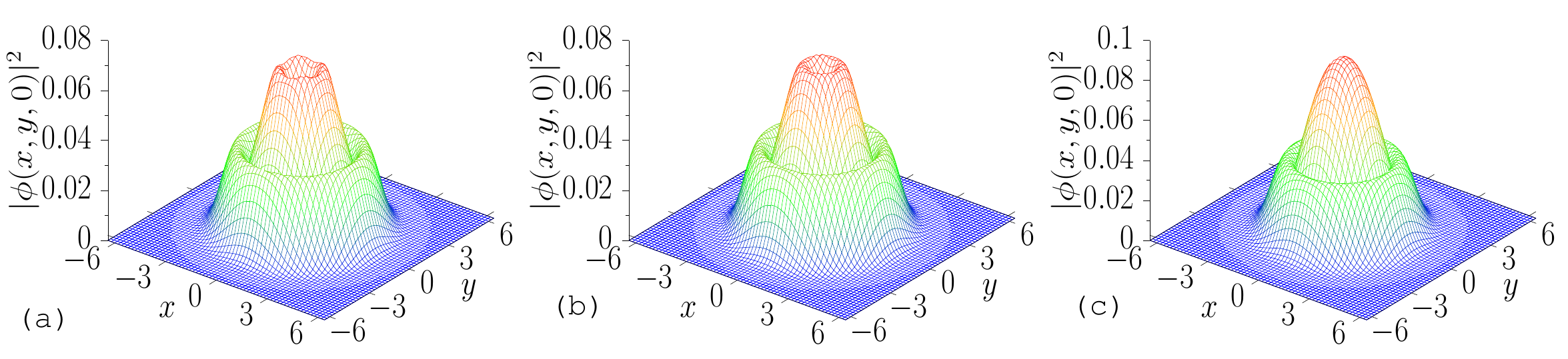}
\end{center}
\caption{(Color online) 
Surface-density plots for the pure-dipolar ($a_{ij}=0$) coupled system $^{168}$Er-$^{164}$Dy, with $\lambda=8$ and fixed  
$N_2=950$ (dysprosium component), shown in three plots as $N_1$ (erbium component) is varied:
(a) $N_1=3000$ (left panel), (b) $N_1 = 2900$ (center panel), and (c) $N_1 = 2000$ (right panel). 
The internal (red) part is dominated by $^{168}$Er, with the surrounding one (green) by $^{164}$Dy.
The biconcave-shaped structure changes to a Gaussian one as we go to a more stable configuration.
}
\label{fig4}
\end{figure*}

In order to illustrate the behavior of the coupled densities, as we go from the parameter region close to the border of stability 
to a more stable configuration, we show three illustrative density plots in Fig.~\ref{fig4} considering the particular case with $\lambda=8$ 
for the mixture $^{168}$Er-$^{164}$Dy, with fixed number of dysprosium atoms $N_2=950$. 
From the panel (a) to panel (c) we are varying the number of erbium atoms from 3000 to 2000.  
We notice that, by going from a more stable region [shown in panel (c), where $N_1=$ 2000] to a region near the 
instability [panel (a), where $N_1=$ 3000] the shape of the density distribution changes from a Gaussian to a 
biconcave format. By approaching very closely the instability region (increasing $N_1$), one can observe the 
fluctuation in the density around the center. 
 
Besides the small mass difference between the atoms, in these cases of pure-dipolar coupled systems, 
the main difference is given by the respective dipolar interactions, which is directly related to the magnetic moment 
dipoles of the components.
While for the $^{168}$Er-$^{164}$Dy we have quite different moment dipoles for the two atomic
components (one is about half of the other), for the mixture $^{164}$Dy-$^{162}$Dy we have about the same
dipolar parameters, with  the inter- and intra-species interactions balancing each other. In view of that, the first
mixture (with erbium) is expected to be more immiscible. This is reflected in the corresponding values of $\eta$
(=0.77 for $^{168}$Er-$^{164}$Dy, and =0.99 for $^{164}$Dy-$^{162}$Dy). 
The results for $\eta$ are consistent with the approximate criterium that one could use for
homogeneous mixtures, where we have $\Delta\approx -0.014$  for $^{168}$Er-$^{164}$Dy, whereas 
$\Delta\approx 0$ for $^{164}$Dy-$^{162}$Dy. However, with the parameter $\eta$ we can have a more 
realistic indication of the partial overlap between the densities.
 
\subsection{Miscibility in non-dipolar coupled systems}
\subsubsection{Role of the trapping aspect ratio}
\begin{figure}[h]
\includegraphics[width=8cm]{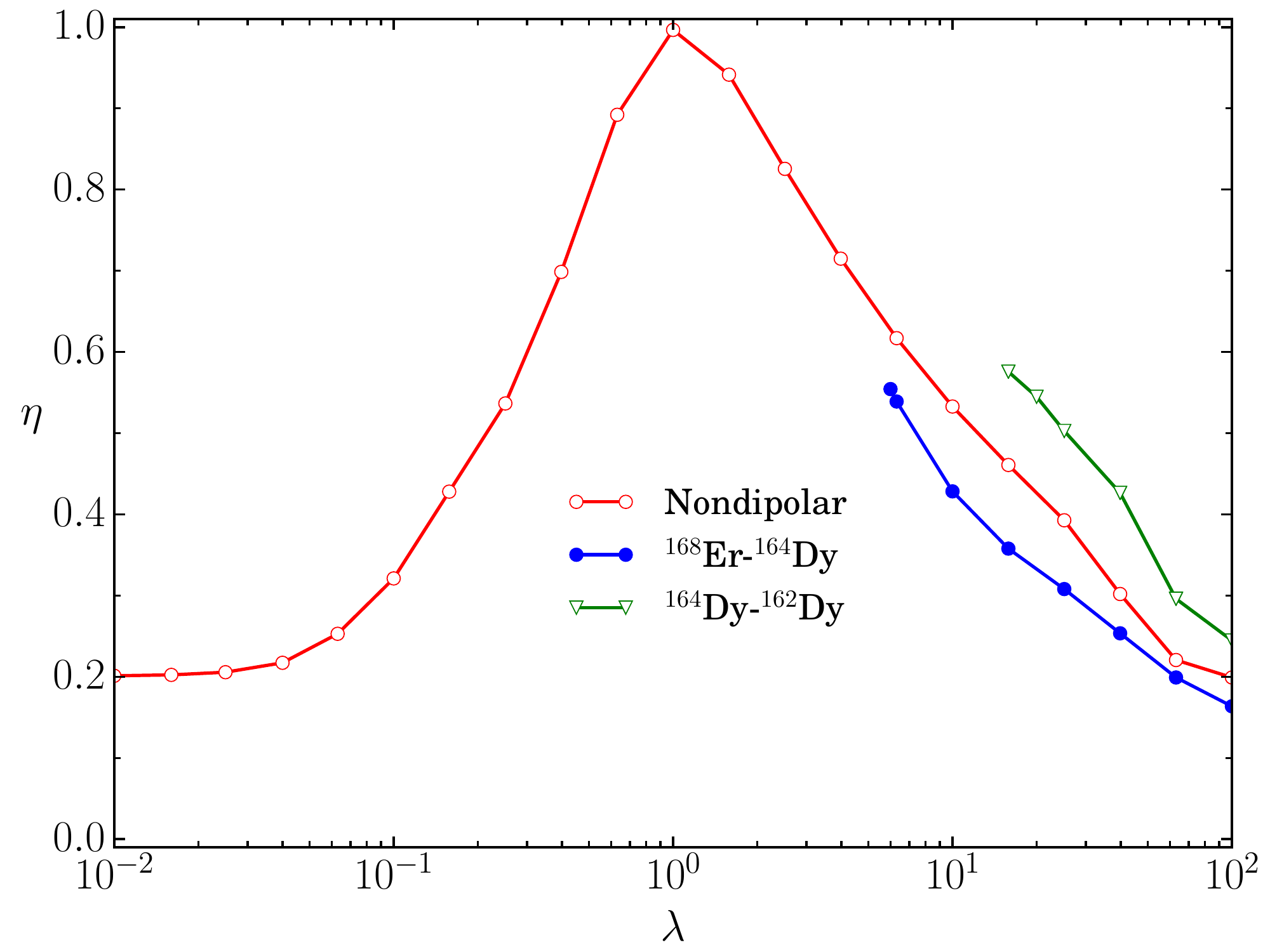}
\caption{(Color online)
The miscibility parameter $\eta$ is shown as a function of $\lambda$, for three configurations, with fixed number of atoms 
($N_1= N_2=6000$) and contact interactions ($a_{11}=a_{22}=40a_0$ and $a_{12}=a_{21}=50a_0$).
For non-dipolar ($a^{(d)}_{ij}=0$) systems $\eta$ is given by a solid-red-circled line.
The solid-blue-with-bullets line, for $^{168}$Er-$^{164}$Dy 
[$a^{(d)}_{11}=66a_0$, $a^{(d)}_{22}=131a_0$ and $a^{(d)}_{12}=a^{(d)}_{21}=94a_0$].
The solid-green-with-triangles, for $^{164}$Dy-$^{162}$Dy 
[$a^{(d)}_{11}=132a_0$, $a^{(d)}_{22}=131a_0$ and $a^{(d)}_{12}=a^{(d)}_{21}=131a_0$].
Stability of the dipolar mixtures restrict the study of $\eta$ to pancake-type traps, having  
$\lambda>6$ for $^{168}$Er-$^{164}$Dy and $\lambda>15$ for $^{164}$Dy-$^{162}$Dy.
}
\label{fig5}
\end{figure}

The effect of the geometry due to the external harmonic trap potential on the phase separation of the mixtures
is investigated in the next, by considering different aspect ratios $\lambda$ for the cases when the
non-linearity is given at least by repulsive two-body interactions. The miscibility parameter, given by the factor $\eta$, 
indicating the amount of mixing in the densities of the two-component, is presented as  a function of $\lambda$ 
in Fig.~\ref{fig5}, for the case that two-body contact interactions are fixed such that 
$a_{11}=a_{22}=40a_0$ and $a_{12}=50a_0$, with fixed number of atoms for both components. 
As discussed when this parameter was defined, a complete overlap 
between the densities implies in $\eta=1$, being zero in the other limit of a complete immiscible system. 
For the given non-zero contact interactions, we observe that a complete miscible  
coupled dipolar system cannot be obained within a stable configuration, as the maximum value verified for the miscibility parameter
is below 0.6 for both dipolar systems and number of atoms that we have considered. However, as we have already pointed out  in the 
previous subsection, a value of $\eta$ near 1 is possible to be obtained in case of complete dipolar systems
(when the contact interactions are set to zero).  

%fig5

The purpose of this subsection is mainly to discuss full non-dipolar systems, with results given in Fig.~\ref{fig5} for non-zero dipolar
interactions (for the $^{168}$Er-$^{164}$Dy, as well as for $^{164}$Dy-$^{162}$Dy). These two kind of mixtures are of interest to show 
how the dipolar interactions can modify the  miscibility behavior of a coupled BEC system. As noticed by the
red-solid line with empty circles given in this figure (with $d_{ij}=0$), for non-dipolar systems we can also obtain stable almost immiscible states
in cigar-type configurations with 
 $\lambda<1$.
With non-zero dipolar parameters, and considering the given contact interactions, 
stable condensates are limited to pancake-type configurations with  $\lambda>6$ for  $^{168}$Er-$^{164}$Dy mixture;
 and  $\lambda>15$ for the $^{164}$Dy-$^{162}$Dy mixture.
As we can see from these results for a non-dipolar coupled system, the immiscibility is more evident when the trap is 
strongly deformed. It can happen for pancake-type traps, as well as for cigar-type traps. However, due to the stability of the 
condensates, it should be more favorable to build immiscible coupled condensates with pancake-type traps. 

\subsubsection{Miscibility in cigar- and pancake-type traps}

%fig6
\begin{figure}[h]
\centering\includegraphics[width=0.99\linewidth]{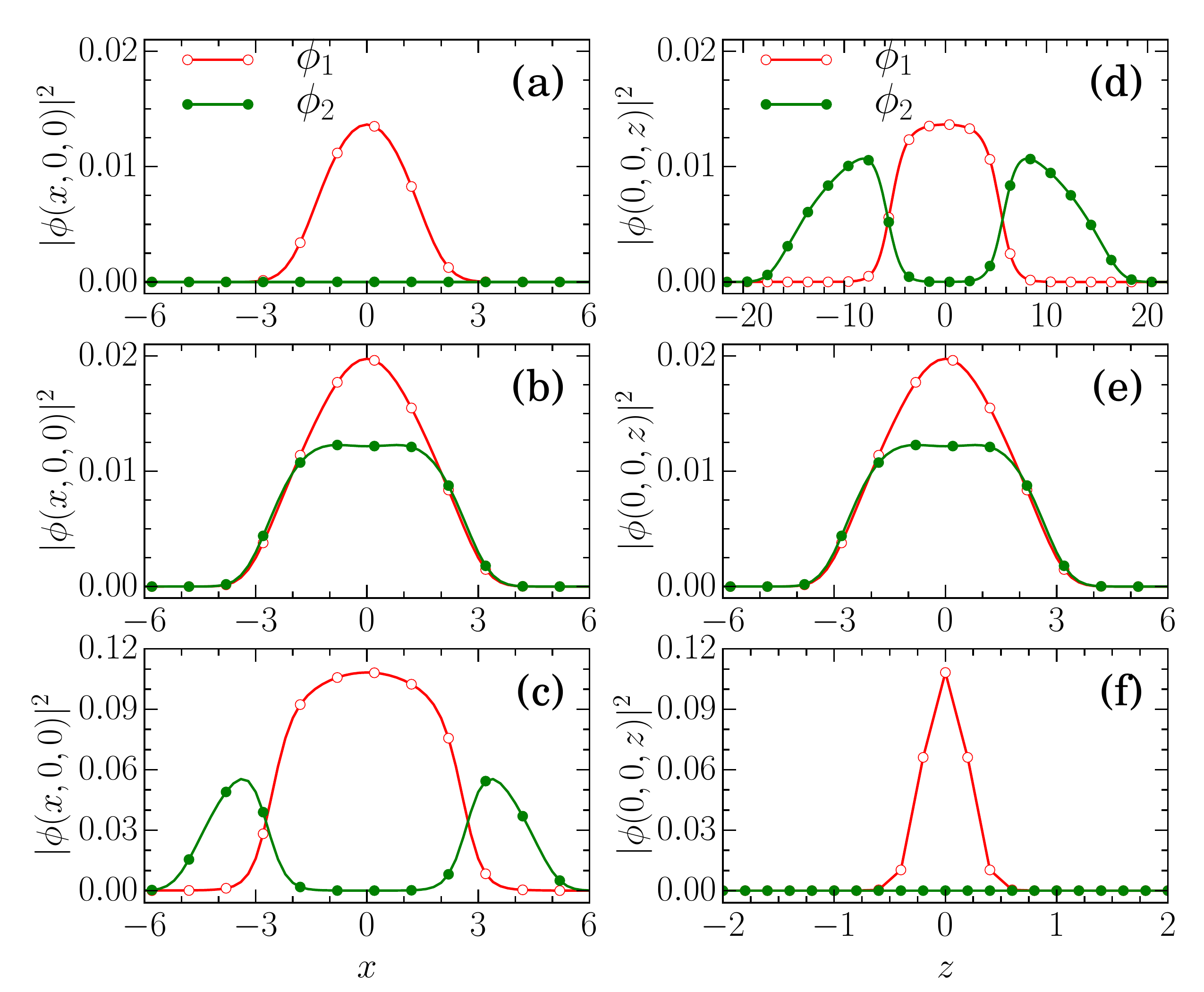}
\caption{ (Color online)
1D plots of the coupled densities for a non-dipolar case, considering three values for the 
aspect ratio: cigar-type $\lambda=0.1$, where  $\eta=0.32$  [panels (a) and (d)]; 
symmetric-case $\lambda=1$, where  $\eta=0.99$ [panels (b) and (e)];  and 
pancake-type $\lambda=20$, where  $\eta=0.43$ [panels (c) and (f)]. The 
 two-body contact parameters and number of atoms $N_i$ are as in Fig.~\ref{fig5}. 
The first component is shown by red lines with empty circles, with the second one shown by green lines 
with solid circles.
In Fig.~\ref{fig7} we add 3D illustrations for the stronger deformed cases, with $\lambda=$0.1 and 20.}
\label{fig6}
\end{figure}

%fig7
\begin{figure}[h]
\begin{center}
\centering\includegraphics[width=0.99\linewidth]{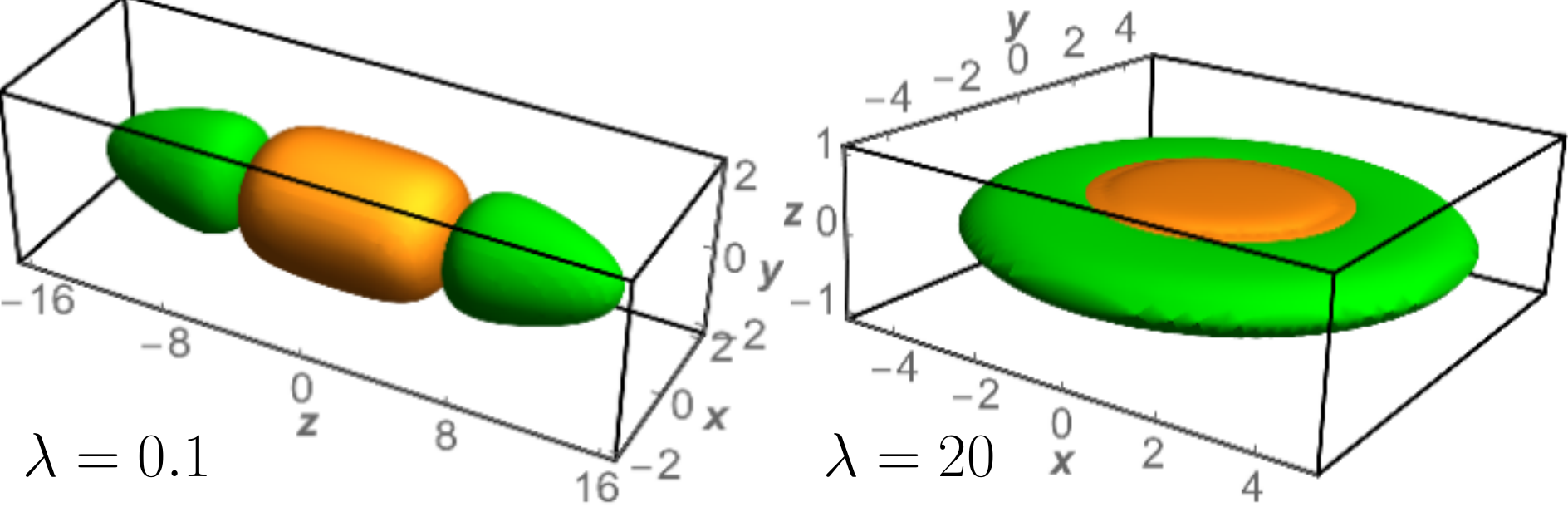}
\end{center}
\caption{(Color online) 
Corresponding to Fig.~\ref{fig6}, we include 3D illustrations 
for the stronger deformed cases, with $\lambda=$0.1 (left panel) and 20 (right panel).
}
\label{fig7}
\end{figure}
For non-dipolar mixtures, we resume our analysis by the 1D density plots given in Fig.~\ref{fig6}, followed by
a corresponding 3D illustration in Fig.~\ref{fig7}. For the parameters we have 
both components with the same number of atoms ($N_1=N_2=6000$), and with repulsive 
two-body interactions such that  $a_{11}=a_{22}=40a_0$ and $\ a_{12}=50a_0$. 
In Fig.~\ref{fig6}, the panels (a), (b) and (c) display the central densities ($y=z=0$) as functions of $x$;
with the panels (d), (e) and (f) showing the densities (for $x=y=0$) as functions of $z$.

From these density plots, we can also infer how well the miscibility parameter $\eta$, given by Eq.~(\ref{eta}), can be quantitatively
used to estimate the miscibility condition of a two-component condensate. 
In the upper panels,  (a) and (d), for a cigar-type trap with $\lambda=0.1$, we have $\eta=0.32$, which is quite well characterizing a 
situation when the component densities have their maximum at distinct positions, with a small overlap between them. Correspondingly, 
we observe similar relations between the value of $\eta=0.43$ and the distinct positions of the extremes for the two components, in the 
case of a pancake-type trap with $\lambda=20$, shown in panels (c) and (f). 
On the other hand, for $\lambda=1$, shown in the panels (b) and (e), we can observe a strong superposition 
of the two components, with their maxima at the same position, with the miscibility parameter close to one 
( $\eta=0.99$).

The two 3D illustrations shown in Fig.~\ref{fig7}, for the strong deformed cases with $\lambda=0.1$ (cigar type) and 
$\lambda=20$ (pancake type)
are also characterizing the corresponding density distributions.

By considering the above discussion on the usefulness of the miscibility parameter $\eta$, in the next density results that will be shown
we rely on the value of this parameter as a relevant quantitative observable for the miscibility analysis and corresponding surface 
visualization of the densities.  As observed, for that we can take $\eta<0.5$ as being almost immiscible, with maxima for the two components 
in well distinct positions. In the other extreme, the system can be considered as almost miscible for $\eta>0.7$. The other intermediate 
cases can be taken as partially miscible systems.

\subsection{Miscibility in coupled BEC with dipolar and contact interactions}
\begin{figure}[htpb]
\begin{center}
\includegraphics[width=0.49\textwidth]{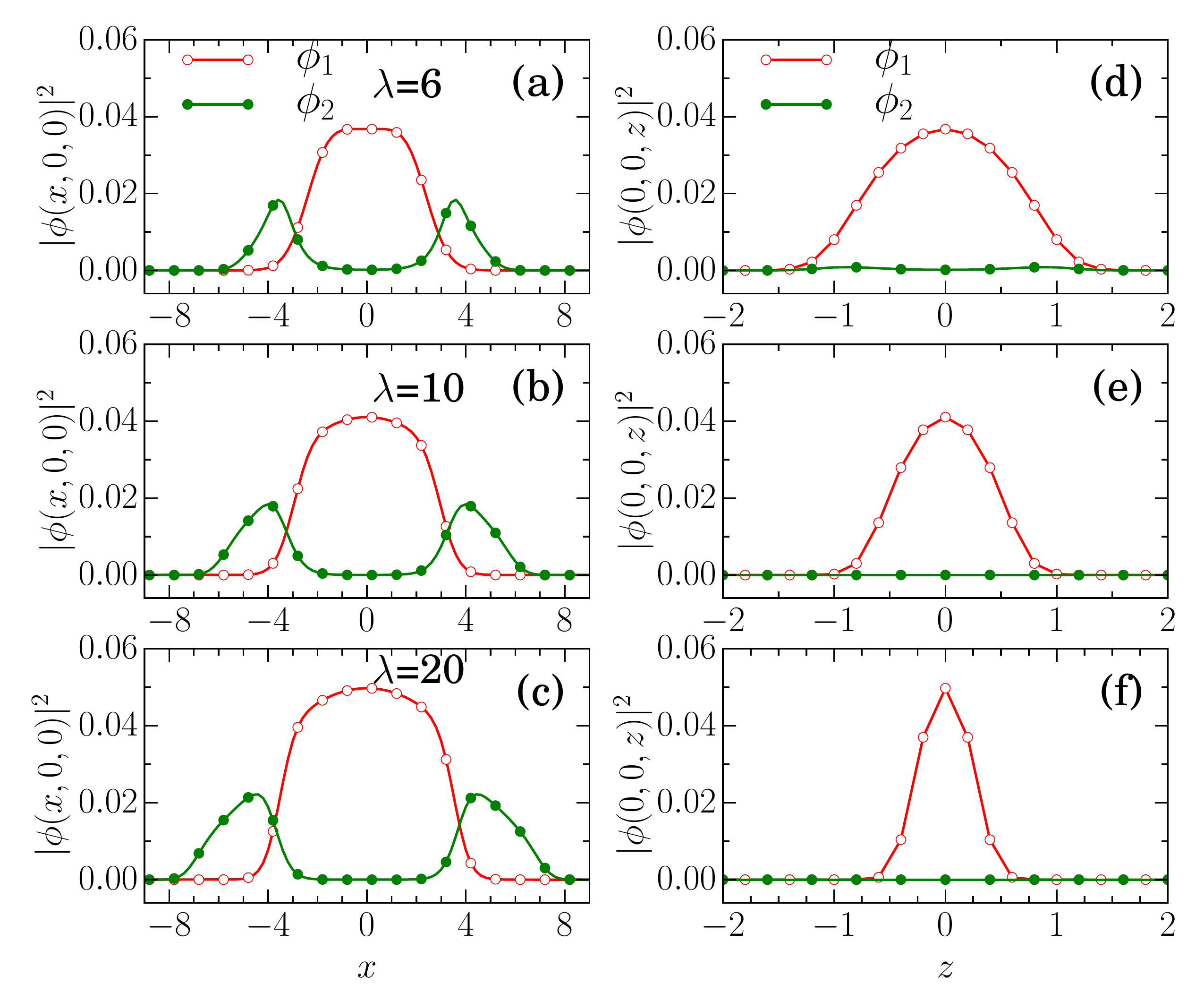}
\end{center}
\caption{(Color online) 
1D plots for the densities of the coupled system $^{168}$Er-$^{164}$Dy system, with both components having the 
same number of atoms, $N_1= N_2=6000$, and subject to non-zero dipolar interactions, given by 
 $a^{(d)}_{11}=66a_0$, $a^{(d)}_{22}=131a_0$ and $a^{(d)}_{12}=a^{(d)}_{21}=94a_0$.
In this case, we have also non-zero intra- and inter-species contact interactions, with
$a_{11}=a_{22}=40a_0$ and $a_{12}=50a_0$. The results are for three different pancake-type traps, with 
$\lambda=$ 6, where $\eta=0.55$ (upper frames); $\lambda=$ 10, where   $\eta=0.43$ (middle frames); and 
$\lambda=$ 20, where $\eta=0.33$ (lower frames). The components are being identified inside the upper panels.}
\label{fig8}
\end{figure}

The miscibility behavior of the coupled condensate in terms of the aspect ratio $\lambda$ is further investigated in
Fig.~\ref{fig8} for dipolar systems in addition to the repulsive two-body contact interactions already used in 
Figs.~\ref{fig6}-\ref{fig7} ($a_{11}=a_{22}=40a_0$, $a_{12}=50a_0$).
In this case, we consider the coupled system $^{168}$Er-$^{164}$Dy, with the 
dipolar parameters given by  $a_{11}^{(d)}=66a_0$, $a_{22}^{(d)}=131a_0$ and $a_{12}^{(d)}=a_{21}^{(d)}=
94a_0$ (respectively, corresponding to $d_{11}=62.8$, $d_{22}=127.8$, $d_{12}=d_{21}=89.5$). By
considering the same atom number for the components $N_1=N_2=6000$, as verified before, stable results are possible only 
for pancake-type trapping potentials with $\lambda>6$.
In general, pancake traps provide strong axial confinement and help to increase the repulsive interaction along radial directions 
which guides to phase separation between the mixtures. In Fig.~\ref{fig8}, by increasing the values for the
aspect ratio (such as $\lambda= $ 6, 10 and 20), we can verify how the structure of the densities for the two components varies
in an immiscible condition. 
The above observations for dipolar and non-dipolar BECs explain their specific characteristics. In particular, 
for erbium-dysprosium coupled dipolar condensate in stable configurations ($\lambda>6$), we are clarifying that 
it presents immiscible density structures, which are more pronounced for larger values of the trap-aspect ratio.

%fig8

\section{Miscibility results - role of the inter-species interaction}
\label{secV}
Besides the trap symmetry, another quite relevant parameter for the miscibility of coupled condensates is
given by the inter-species interaction.
In this case, when considering a system with fixed dipolarity, the appropriate parameter, which can be tuned 
via Feshbach resonance techniques~\cite{2010-chin}, is given by the inter-species contact interaction.
Therefore, our aim in the following analysis is to characterize the role of the inter-species scattering length 
$a_{12}$ for the miscibility. For that, we assume both species have the same intra-species scattering lengths,
$a_{11}=a_{22}=40a_0$, choosing a particular large pancake-type trap with $\lambda=20$ and 
fixed number of atoms for both components, $N_1=N_2=6000$. 
These parameters are dictated by the previous stability analysis, looking for stronger characterization 
of miscibility properties. 

\begin{figure}[htpb]
\begin{center}
\includegraphics[width=0.4\textwidth]{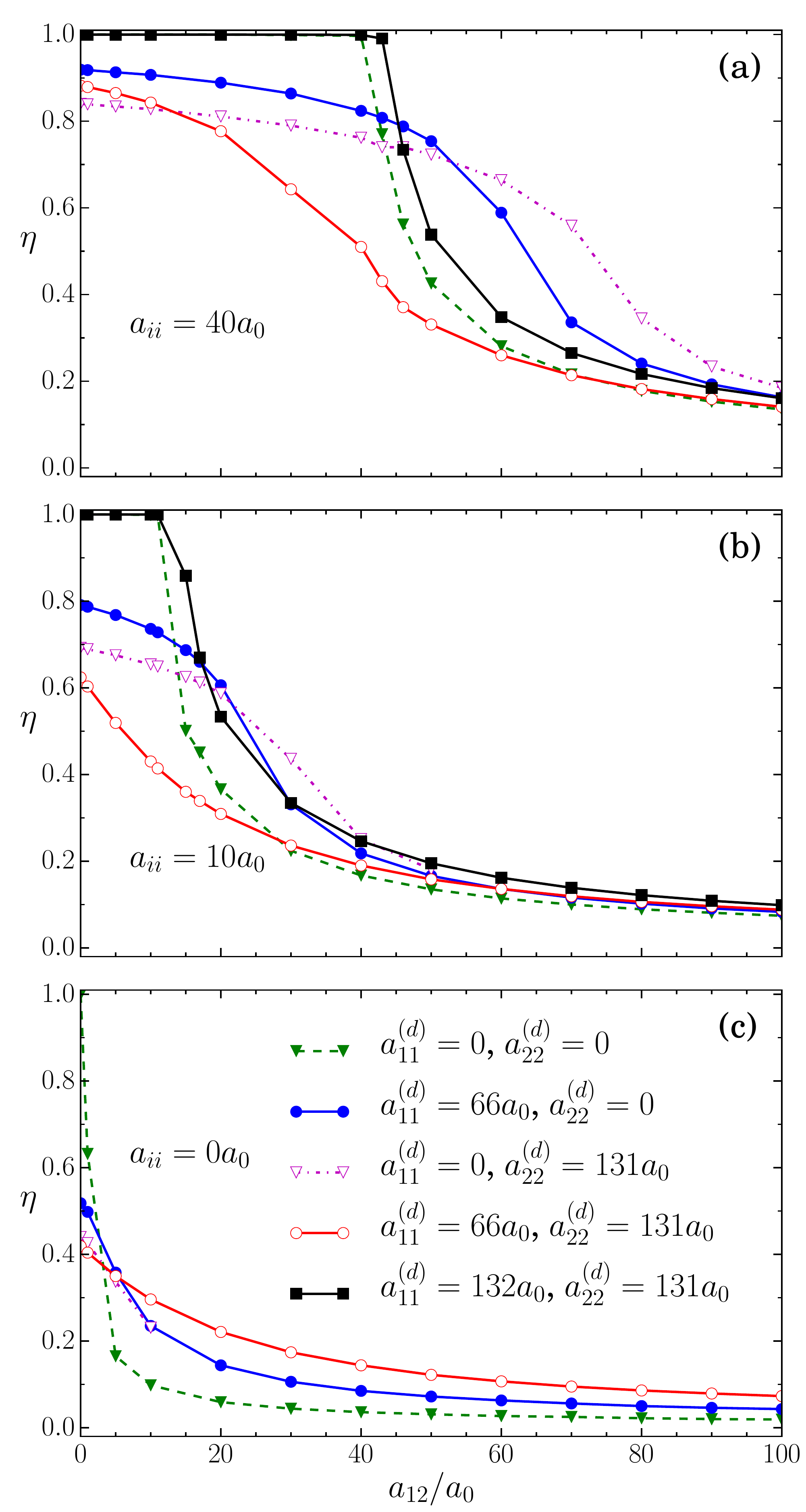}
\caption{(Color online)
The miscibility $\eta$ is shown as a function of 
$a_{12}$ in three frames, with fixed intra-species contact interactions, 
$a_{11}=a_{22}=$  $40a_0$ (a), $10a_0$ (b) and $0$ (c). The aspect ratio is $\lambda=20$,
with number of atoms $N_1=N_2=6000$.  
The dipolar parameters used for the curves shown in each panel are given inside the lower frame.
When the system turns out to be unstable, the lines are interrupted or no-results shown.
} 
\label{fig9}
\end{center}
\end{figure}

The general behavior for the miscibility can be verified by the parameter $\eta$, which is
shown in Fig.~\ref{fig9} as a function of the ratio $a_{12}/a_{0}$, considering three panels with different values for 
$a_{ii}/a_0=$40, 10 and 0.  The parameter $\eta$, for the overlap between the two-component densities, is represented 
by considering five cases as indicated inside the frames, where in the lower frame we have the dipolar parameters valid 
for the three frames.  When both components are dipolar, the mixture $^{168}$Er-$^{164}$Dy is represented by the red 
curves with empty circles, with the mixture $^{164}$Dy-$^{162}$Dy represented by black curves with solid squares.
When both components are non-dipolar, we show with green-curves with triangles. 
When only a single component has dipolar interactions, we have two cases, represented with blue curves with 
filled circles ($a_{11}^{(d)}=66a_0, a_{22}^{(d)}=0, $)
and magenta line with empty triangles ($a_{11}^{(d)}=131a_0, a_{11}^{(d)}=0 $).

%fig9

We should point out that  the miscibility is going down slowly (increasing the immiscibility of the mixture)
to an approximate constant value for $a_{12}\to \infty$.  Even in the non-dipolar case: with $a_{ii}=$40, 10 and 0,
we have $\eta\sim$ 0.14, 0.10 and 0.04, respectively. This can be understood from the residual mixing of the 
two-component wave-functions near the unitary limit of the inter-species (when $a_{12}\to \infty$). 
In this limit for the inter-species, we should notice that we have the well-known Efimov effect~\cite{efimov}, 
with increasing number of three-body bound and resonant states mixing the two
component system. Already observed in cold-atom laboratories~\cite{kraemer-2006}, this a pure quantum effect 
that occurs near zero two-body binding, when the
effective potential goes as the inverse-square of the distance, and we have a long extension of the corresponding 
two-body wave functions.
The role of this effect on coupled systems deserves further analysis in experiments, in particular when varying 
the mass-ratio of the mixture, as the resonant states have well-known theoretical predictions~\cite{3body} also
for mass-asymmetric mixtures.

Next, we can examine the role of the magnetic dipolar interactions in the miscibility of the two components. First, we should
remind that these interactions between the magnetic dipoles are long-range ones, which go as the third power of the distance
between the dipoles. They should act more effectively when the components are close together, but have residual
effects due to long range behavior, which makes the miscibility be reduced to a non-zero constant value for $a_{12}\to \infty$,
in a way similar to the non-dipolar case in this limit, such that we have combined effects of two long-range interactions.
In the other extreme, we can see the role of dipolar interactions in the miscibility by looking the 
region where $a_{12}=0$ in the lower panel. 
When we have no inter-species dipolar interactions (with only one of the species having magnetic moment) there is 
no coupling between the systems, such that the overlap between the densities is partial. 
 One of the densities is given by a linear trapped equation; with the other given by a non-linear equation
 (confined by the same harmonic trap, but additionally having the repulsive dipolar interaction). 
 As larger is the effect due to repulsive dipolar interactions, less miscible is the system.
When both intra-species dipolar interaction are switched on, we have also the corresponding inter-species parameter 
coupling the system, with the net repulsive effect being averaged out.
 
Another aspect verified in the panels of Fig.~\ref{fig9} refers to the stability of the system, which can be observed for this
specific case that we have a pancake-shaped trap with $\lambda=20$ and fixed number of atoms $N_i=6000$.
When we have single dipolar interaction with larger strength as $a^{(d)}_{22}=131a_0$, the stability can happen only for 
some limited values of the contact interactions, as indicated in the three panels by the maximum values obtained for 
$a_{12}$. For $a_{ii}=0$ (lower panel), $a_{12}\le 10 a_0$;  for $a_{ii}= 10a_0$ we obtain $a_{12}\le 50 a_0$; 
with no limit in  $a_{12}$ when $a_{ii}= 40 a_0$.
These upper limits in $a_{12}$ for the stability result from the specific trap conditions and number of atoms we are considering, 
as one could trace, approximately, from the stability results shown in Fig.~\ref{fig1}.  Therefore, by reducing the number of 
atoms for the component 2, we could increase the upper limits for $a_{12}$. 

In general, from Fig.~\ref{fig9} one can observe that the immiscibility will significantly increase by increasing $a_{12}$ in 
all the verified situations.  Also, by taking a fixed value for this inter-species contact interaction with $a_{12}>a_{ii}$, 
from the three panels one can verify that by decreasing the intra-species contact interactions $a_{11}=a_{22}$, the immiscibility 
of the system increases ($\eta$ decreases).
 
Density profiles of the coupled system are represented in the next Figs.~\ref{fig10} - \ref{fig13}, where we show results 
considering three values of the two-body inter-species scattering lengths, with the corresponding intra-species contact 
interactions $a_{ii}$ kept fixed at $40a_0$. In all these cases, in order to facilitate the comparison of the results, we assume 
the same three values for $a_{12}$ ($= 10a_0$, $43 a_0$ and $60 a_0$) and maintain the trap-aspect ratio fixed to a 
pancake-type, $\lambda=20$, with the number of particles given by $N_1=N_2=6000$, as in the plotted results of Fig.~\ref{fig9}.
For a better quantitative comparison of the results, we present the densities in 1D plots, as functions of $x$ (left panels)
and $z$ (right panels). In Fig.~\ref{fig10}, we have the case of non-dipolar systems; in Fig.~\ref{fig11} when considering  
just one of the components being dipolar. In Figs.~\ref{fig12} and \ref{fig13}, we consider two cases, when both components 
are dipolar, with parameters corresponding to the mixtures $^{168}$Er-$^{164}$Dy and  $^{164}$Dy-$^{162}$Dy, respectively.

%fig10
\begin{figure}[htpb]
\begin{center}
\includegraphics[width=0.49\textwidth]{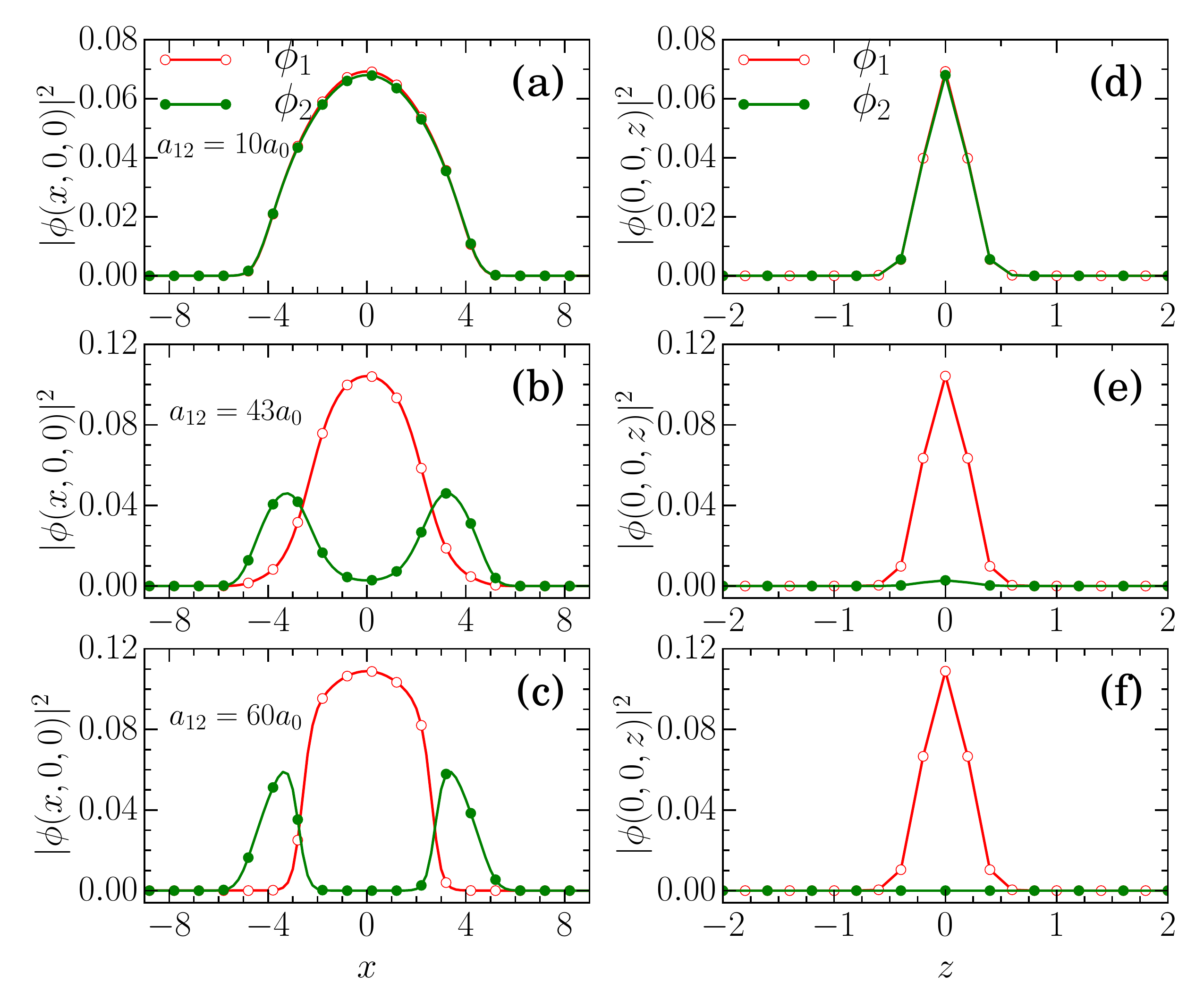}
\end{center}
\caption{(Color online) 1D plots of the densities for a non-dipolar system, by varying the inter-species two-body interactions, 
with $a_{12} =10 a_0$, where $\eta=1.0$ (upper frames);  $a_{12} =43 a_0$, where $\eta=0.77$ (middle frames); and 
$a_{12} =60 a_0$, where $\eta=0.28$ (lower frames).  
The other scattering lengths are fixed, with $a_{11}=a_{22}=40a_0$. The trap aspect ratio is $\lambda$ = 20,  
and the number of condensed  atoms is the same, $N_1= N_2=6000$, for both $^{168}$Er  and $^{164}$Dy components.
The components are being identified inside the upper panels.}
\label{fig10}
\end{figure}

%fig11
\begin{figure}[htpb]
\begin{center}
\includegraphics[width=0.49\textwidth]{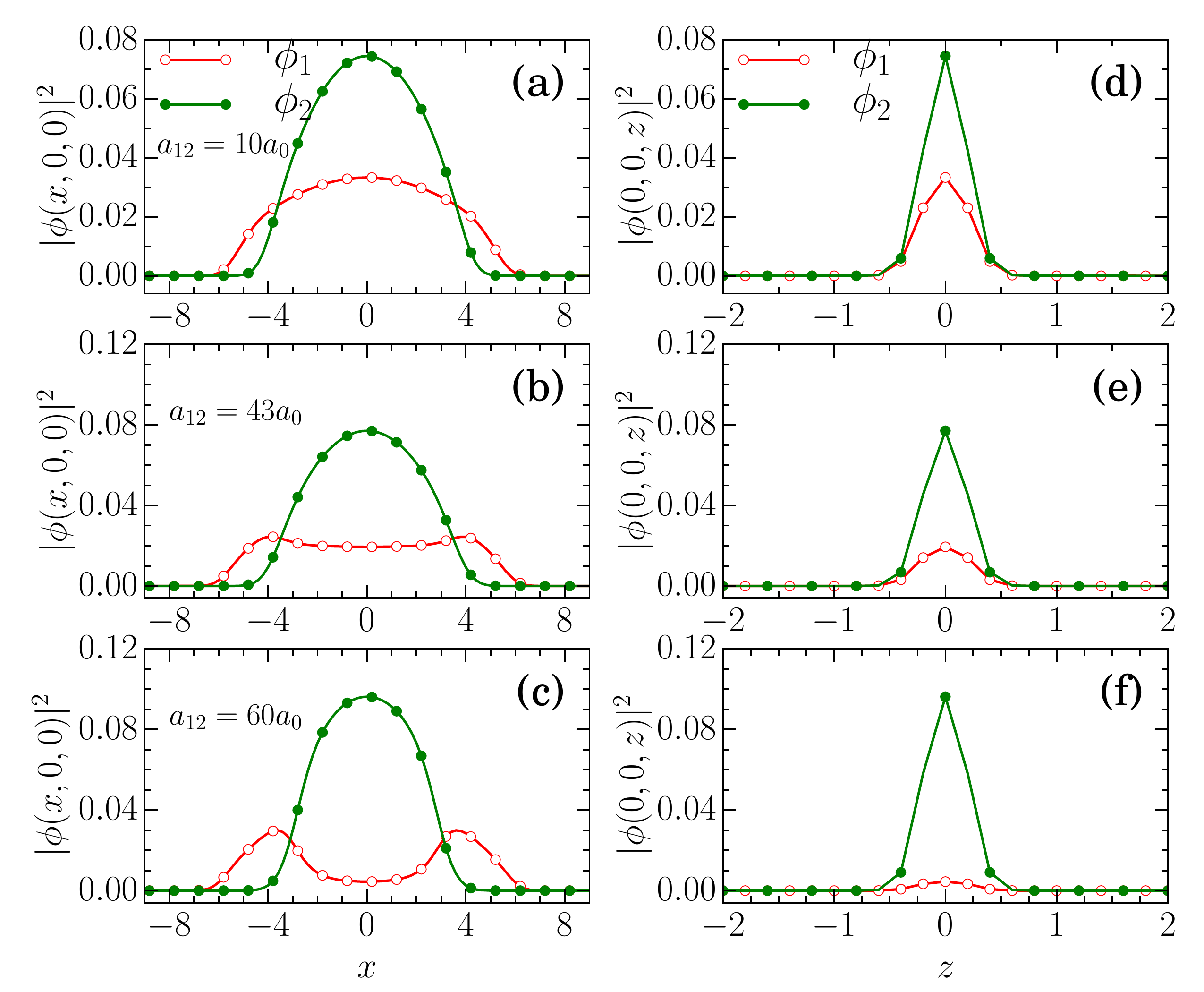}
\end{center}
\caption{(Color online) 
1D plots of the densities, with component $1$ being dipolar and component 2 non-dipolar, such that  
$a^{(d)}_{11}=66a_0$, $a^{(d)}_{22}=a^{(d)}_{12}=a^{(d)}_{21}=0$,
by considering three different inter-species two-body interactions:  
$a_{12} =10 a_0$, where $\eta=0.91$ (upper frames);  $a_{12} =43 a_0$,
where $\eta=0.81$ (middle frames); and $a_{12} =60 a_0$, where $\eta=0.59$ (lower frames). 
As in Fig.~\ref{fig10}, $\lambda$ = 20, $N_1= N_2=6000$, and the intra-species contact
interactions are fixed with $a_{11}=a_{22}=40a_0$.
The components are being identified inside the upper panels. 
}
\label{fig11}
\end{figure}

%fig12
\begin{figure}[!ht]
\centering\includegraphics[width=0.5\textwidth]{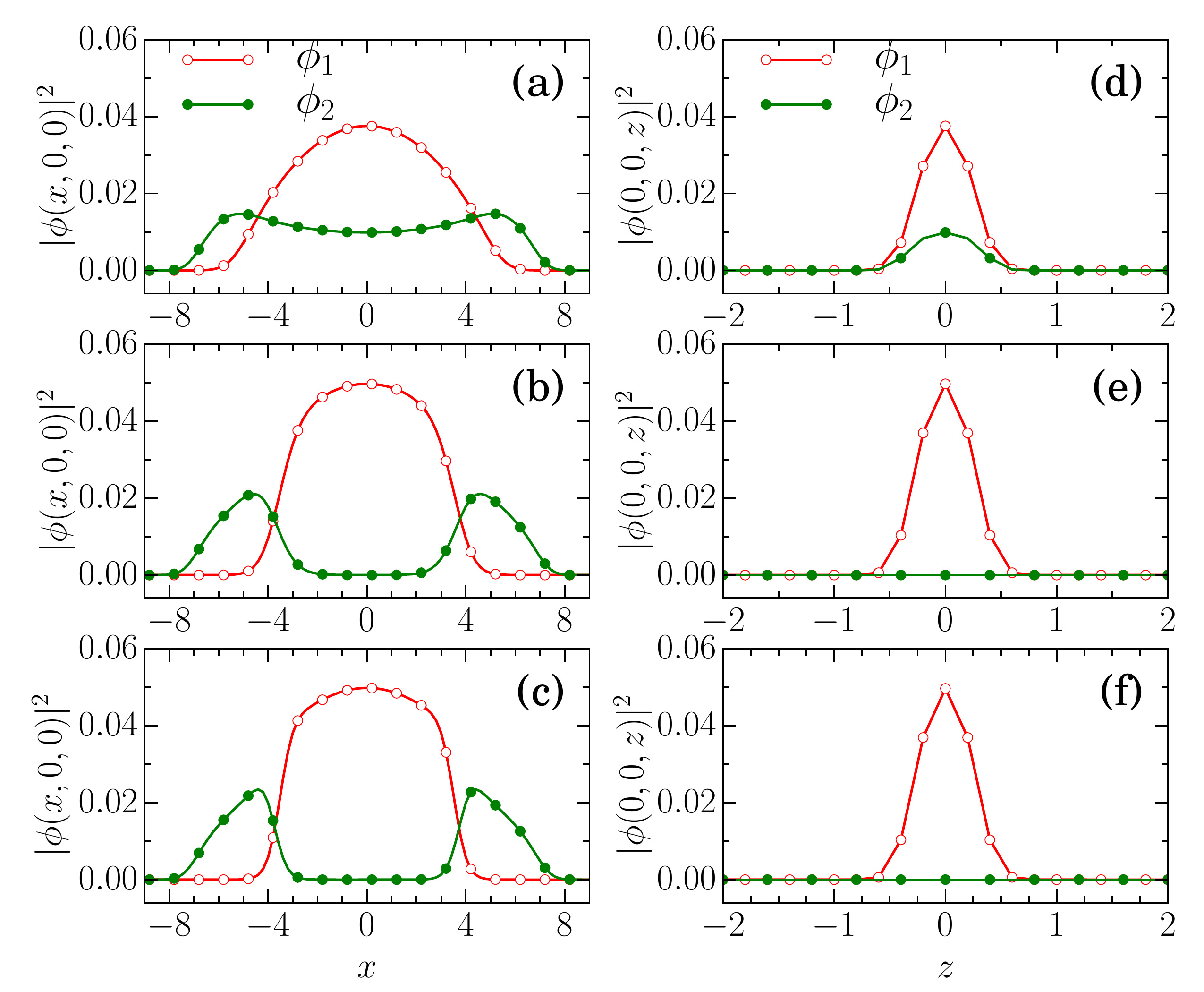}
\caption{(Color online) 
1D plots of the densities, for the case that both components have non-zero dipolar interactions, 
with $a^{(d)}_{11}=66a_0$, $a^{(d)}_{22}=131a_0$ and $a^{(d)}_{12}=a^{(d)}_{21}=94a_0$,
considering three different inter-species two-body contact interactions, such that
$a_{12} =10 a_0$, where $\eta=0.84$ (upper frames); $a_{12} =43 a_0$, where $\eta=0.41$ 
(middle frames); and  $a_{12} =60 a_0$, where $\eta=0.26$ (lower frames).
As in Figs.~\ref{fig10} and \ref{fig11}, $\lambda$ = 20, $N_1= N_2=6000$, and the intra-species 
scattering lengths are fixed such that $a_{11}=a_{22}=40a_0$. 
The components are being identified inside the upper panels.
}
\label{fig12}
\end{figure} 

%fig13
\begin{figure}[htpb]
\begin{center}
\includegraphics[width=0.49\textwidth]{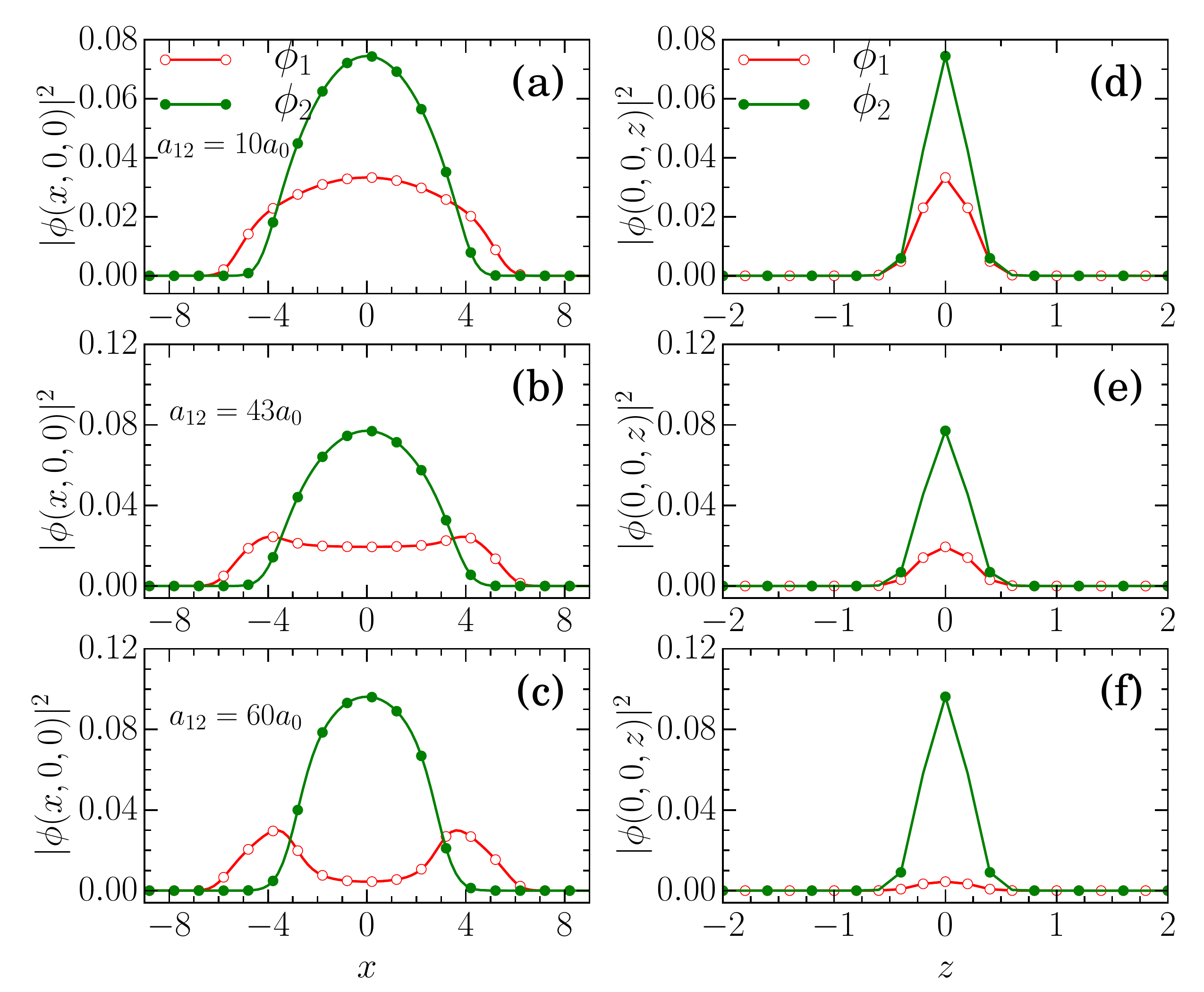}
\end{center}
\caption{(Color online)  1D plots of the densities for the case of $^{164}$Dy-$^{162}$Dy mixture,
with $a^{(d)}_{11}=132a_0$, $a^{(d)}_{22}=131a_0$ and $a^{(d)}_{12}=a^{(d)}_{21}=131a_0$,
considering different inter-species two-body contact interactions, such that
$a_{12} =10 a_0$, where $\eta=0.99$ (upper frames); $a_{12} =43 a_0$, where $\eta=0.91$ 
(middle frames); and  $a_{12} =60 a_0$, where $\eta=0.354$ (lower frames).
As in Fig.~\ref{fig12}, $\lambda$ = 20, $N_1= N_2=6000$, and the intra-species 
scattering lengths are fixed such that $a_{11}=a_{22}=40a_0$. 
The components are being identified inside the upper panels.
} 
\label{fig13}
\end{figure}

First, by considering non-dipolar BECs, in Fig.~\ref{fig10}, with three values for the inter-species scattering length, 
($a_{12}$ varying between $10$ and $60a_0$, with $a_{11}=a_{22}=40a_0$).
The transition is expected to occur close to $a_{12}\approx a_{11}$ ($m_1\approx m_2$). 
This is consistent with the results shown in this figure, where we can verify that for  $a_{12}>a_{11}$ there is already 
a clear separation between the two species, with $\eta=0.28$ for $a_{12}=60a_0$. It is also shown that 
for $a_{12}=10a_0$ the mixture is completely miscible ($\eta=1$), having a sudden transition for $a_{12}=43a_0$ 
(just above $a_{11}=40a_0$) to a partial immiscible state. By further increasing $a_{12}$, an almost complete 
immiscible state is reached (for $a_{12}=60a_0$, $\eta=0.28$) . 

 Next, in Fig.~\ref{fig11}, we study the case where only the component 1 is dipolar. 
 By comparing with the non-dipolar case, we can observe the effect of the dipolar interaction in breaking
 the sudden transition verified when increasing the inter-species contact interaction, such that the 
 transition  is much softer than the one shown for the non-dipolar case.  
 
In order to observe the effect in the miscibility when changing the inter-species scattering length for the
cases that both components are dipolar, we present 1D  plots for the densities in  Figs.~\ref{fig12} and \ref{fig13}. 
For the kind of mixture considered in \ref{fig12} , the $^{168}$Er-$^{164}$Dy,
we notice that the system is partially miscible even at $a_{12}<a_{11}$, becoming almost immiscible for 
$a_{12} > 30a_0$. However, in case of $^{164}$Dy-$^{162}$Dy mixture, we observe that the system is 
almost miscible for small inter-species scattering lengths, becoming immiscible as we increase the inter-species
scattering length. Therefore, we clearly noticed from these two figures, the immiscibility increasing when the 
inter-species scattering length become dominant in respect to the dipolar interactions. 

\section{Summary and Conclusion}
\label{secVI}
In the present work, our approach was to evidence the miscibility properties of a coupled condensate 
with two different species of atoms, having contact or dipolar pairwise interaction between them.
The confining potential for the coupled system is assumed having harmonic shape, being symmetric 
 in a plane with an asymmetry along the third axis by a given trap-aspect parameter $\lambda$. 
Within our 3D Gross-Pitaevskii-type formalism, the nonlinearity is originated from two-body 
contact interactions as well as from dipolar interactions. First, we did an investigation related to the
stability of dipolar coupled systems, as one varies the trap-aspect ratio and the number of atoms 
of both species. This study help us to define stable configurations (and corresponding parameters) 
that are more appropriate to study the miscibility. In view of that, for realistic number of atoms in a mixed BEC
system, we found necessary to consider pancake-type configurations ($\lambda>1$) for the coupled 
condensates, when the dominant nonlinear interaction is dipolar. 
In order to be more complete on the characterization of miscibility, we have extended our study to 
non-dipolar systems in cigar-type configurations, where it was possible to point out strong immiscibility 
for the coupled system.

Motivated by recent experimental studies with dipolar systems, when considering dipolar atomic systems,
we focus our study in the two coupled mixtures given by $^{168}$Er-$^{164}$Dy and $^{164}$Dy-$^{162}$Dy. 
In this respect, our theoretical predictions can be helpful in verifying miscibility properties in on-going experiments 
under different anisotropic trap configurations. 
When considering non-dipolar coupled systems, our predictions on the miscibility can be more generally be 
investigated, considering several non-dipolar atomic mixtures, as there are no stability requirements to 
constrain the values of the positive contact parameters. 
For the sake of generalization, we have also examined the case when 
only one of the species is dipolar.

The stability properties had to be stablished for the binary dipolar BEC system that we have investigated,
before studying the corresponding miscibility behavior, considering the fraction number of atoms for each 
species as functions of the trap-aspect ratio $\lambda$. We assumed both species confined in traps 
having the same aspect ratio.

For the study of miscibility, we start by extending an approach for the critical conditions of 
homogeneous coupled systems confined in hard-wall barriers, in order to point out how to vary the contact 
and dipolar parameters for a transition from miscible to immiscible configurations. 
It was observed that the critical MIT conditions remain unaffected by the dipolar interactions, once all
the parameters of the previous definition are rescaled by incorporating the dipolar ones.

In order to measure the miscibility of a more general confined coupled system, a relevant parameter $\eta$ was 
defined in terms of the two-component densities. This parameter is found appropriate for a quantitative estimate 
of the overlap between the two densities of the coupled system, which is equal to one for a complete miscible system, 
reducing to zero when there is no overlap between the densities (an immiscible configuration). 
It is worthwhile to comment that in our definition of $\eta$ we are integrating the square-root of the product of 
the component densities. This was done in order  to keep some similarity with the definition given in 
Ref.~\cite{phase-param}, where the overlap of the wave-functions was used. Therefore, one could use 
$\eta^2$ for a more realistic picture of the overlap region of the densities. For a general system, this parameter is 
shown to be adequate to verify the miscibility of a coupled system than the usually simplified criterium obtained for 
homogeneous systems from energy consideration, where the kinetic energy is ignored. 

By studying the miscibility for pure-dipolar coupled system (zero two-body contact interactions), we first compare
the properties of the two mixtures given by  $^{168}$Er-$^{164}$Dy and $^{164}$Dy-$^{162}$Dy. Besides the 
small mass difference between the atoms in both this two mixtures, one should noticed that the main difference in
their respective dipolar interactions  $a_{ij}^{(d)}$ is due to the differences in the magnetic moment dipole of erbium 
and dysprosium atoms. 
As verified, the two mixtures have quite different miscibility behavior, with $^{164}$Dy-$^{162}$Dy being completely 
miscible ($\eta=0.99$) and $^{168}$Er-$^{164}$Dy partially miscible ($\eta=0.77$), when we fix to the same values 
the other parameters (trap-aspect ratio and number of atoms). Such behavior is clearly due to the inter-species
dipolar strength in comparison with the intra-species one.

In this pure-dipolar case, we are also pointing out some non-trivial biconcave-shaped structures, with local minimum states 
for both components of the coupled system and with the emergence of density oscillations (manifested by a few peaks), 
when the system is near the stability border. This behavior is verified for both coupled mixtures that we have examined,
with no direct relation with the miscibility of the species.
Already reported in Ref.~\cite{martin-2012}, this kind of nontrivial biconcave configuration has also being 
observed for single component dipolar systems, with the density fluctuations attributed to roton mode of the
condensates. 

The role of the trap aspect ratio and inter-species contact interaction for the miscible-immiscible 
phase transition was studied for different configurations, from non-dipolar to pure dipolar systems.
For dipolar cases, we verify that pancake trap provides strong repulsive interactions along the radial direction, 
leading to clear defined immiscible phases between the components, which are basically controlled by
the trap-aspect ratio.
We have also verified that immiscible states can be well characterized for cigar-type traps, for
non-dipolar systems with repulsive two-body contact interactions.  In this case,
our results were restricted only to a non-dipolar case, in view of stability conditions when 
considering enough large number of atoms for the coupled system chosen in the present study. 

Finally, the present results on the miscibility are expected to be quite relevant in studies with dipolar
and non-dipolar coupled systems, in order to fix parameters in possible going on experimental investigations. 
Within this perspective, when considering dipolar systems, we choose atomic mixtures with large repulsive 
dipolar strengths that are being investigated in BEC laboratories. 
In a possible straightforward extension of this work, we could alter the confining conditions of both condensates,
with their center being separated by some distance, or by using different aspect ratios.
As another perspective for future developments, we can mention studies of systems under rotations, where rich 
vortex structures will emerge, by following recent interest in the subject that can be traced 
from \cite{Kishor2016} and references therein.

\acknowledgments
RKK acknowledges the financial support from FAPESP of Brazil (Contract number 2014/01668-8). 
The work of PM forms a part of Science \& Engineering Research Board, Department of Science \& Technology 
(India) sponsored research project (No. EMR/2014/000644) and FAPESP (Brazil). 
AG and LT thank CAPES, CNPq and FAPESP of Brazil for partial support.


\begin{thebibliography}{99}

\bibitem{1995-bec-exp}
M. H. Anderson, J. R. Ensher, M. R. Matthews, C. E. Wieman, and E. A. Cornell, Observation of Bose-Einstein
condensation in a dilute atomic vapor, Science {\bf 269}, 198-201 (1995);
K.B. Davis, M.-O. Mewes, M.R. Andrews, N.J. van Druten, D.S. Durfee, D.M. Kurn, and W. Ketterle (1995), Bose-Einstein condensation in a gas of sodium atoms, \prl {\bf 75}, 3969-3973 (1995);
C. C. Bradley, C. A. Sackett, J. J. Tollett, and R. G. Hulet, 
Evidence of Bose-Einstein Condensation in an Atomic Gas with Attractive Interactions,
\prl {\bf 75}, 1687 (1995).

\bibitem{2002-08-becrev}
J.R. Anglin and W. Ketterle, Bose-Einstein condensation of atomic gases,
Nature {\bf 416}, 211-218 (2002);
S. Chu, Cold atoms and quantum control,
Nature {\bf 416}, 206-210 (2002);
I. Bloch, J. Dalibard, and W. Zwerger, Many-body physics with ultracold gases, 
Rev. Mod. Phys. 80, 885 (2008).

\bibitem{2017-proukakis}
N.P. Proukakis, D.W. Snoke, and P.B. Littlewood (Eds), 
{\it Universal Themes of Bose-Einstein Condensation}, Cambridge University Press, 2017.

\bibitem{2012-block} I. Bloch, J. Dalibard, and S. Nascimb\`ene,
Quantum simulations with ultracold quantum gases,
Nature Physics {\bf 8}, 267-276 (2012).

\bibitem{2010-chin}
C. Chin, R. Grimm, P. Julienne, and E. Tiesinga, Feshbach resonances in ultracold gases,
Rev. Mod. Phys. {\bf 82}, 1225 (2010).

\bibitem{bec-works}
M. Greiner, C. A. Regal and D. S. Jin, Emergence of a molecular Bose-Einstein condensate from a Fermi gas, 
Nature {\bf 426}, 537-540 (2003); 
I. Bloch, Ultracold quantum gases in optical lattices, Nature Phys. {\bf 1}, 23-30, (2005);
I. B. Spielman, W. D. Phillips and J. V. Porto, Mott-insulator transition in a two-dimensional atomic Bose gas, 
\prl {\bf 98}, 080404 (2007);
H.P. B\"uchler, E. Demler, M. Lukin, A. Micheli, N. Prokof'ev, G. Pupillo, and P. Zoller, 
Strongly correlated 2D quantum phases with cold polar molecules: controlling the shape of the interaction potential, 
\prl {\bf 98}, 060404 (2007);
R. J\"ordens, N. Strohmaier, K. G\"unter, H. Moritz and T. Esslinger, A Mott insulator of fermionic atoms 
in an optical lattice, Nature {\bf 455}, 204-207 (2008); 
S. Kraft, F. Vogt, O. Appel, F. Riehle, and U. Sterr, Bose-Einstein condensation of alkaline 
earth atoms: $^{40}$Ca, \prl {\bf 103}, 130401 (2009);
P. Schauß, M. Cheneau, M. Endres, T. Fukuhara, S. Hild, A. Omran, T. Pohl, C. Gross, S. Kuhr, and I. Bloch, 
Observation of spatially ordered structures in a two-dimensional Rydberg gas, Nature {\bf 491}, 87-91 (2012).

\bibitem{Myatt1997}
C. J. Myatt, E. A. Burt, R. W. Ghrist, E. A. Cornell, and C. E. Wieman, Production of two overlapping 
Bose-Einstein condensates by sympathetic cooling, \prl {\bf 78}, 586 (1997)

\bibitem{2004-chui} S. Chui and V. Ryzhov, Collapse transition in mixtures of bosons and fermions, 
Phys. Rev. A. {\bf 69} 043607 (2004).

\bibitem{1998-Hall-Phase-sep} 
D. S. Hall, M. R. Matthews, J. R. Ensher, C. E. Wieman, and E. A. Cornell, 
Dynamics of component separation in a binary mixture of Bose-Einstein condensates, 
\prl {\bf 81}, 1539 (1998); 

\bibitem{2010-Tojo-phasecontrol}
S. Tojo, Y. Taguchi, Y. Masuyama, T. Hayashi, H. Saito, and T. Hirano, Controlling phase separation of binary 
Bose-Einstein condensates via mixed-spin-channel Feshbach resonance, \pra {\bf 82}, 033609 (2010);

\bibitem{2008-Papp-Tune-Mis}
S. B. Papp, J. M. Pino, and C. E. Wieman, 
Tunable Miscibility in a Dual-Species Bose-Einstein Condensate, \prl {\bf 101}, 040402 (2008)

\bibitem{K-Rb-Cs}
G. Ferrari, M. Inguscio, W. Jastrzebski, G. Modugno, G. Roati, and A. Simoni, 
Collisional properties of ultracold K-Rb mixtures, \prl {\bf 89}, 053202 (2002);
G. Thalhammer, G. Barontini, L. De Sarlo, J. Catani, F. Minardi, and M. Inguscio, 
Double species Bose-Einstein condensate with tunable interspecies interactions, \prl {\bf 100}, 210402 (2008); 
D. J. McCarron, H. W. Cho, D. L. Jenkin, M. P. K\"oppinger, and S. L. Cornish, 
Dual-species Bose-Einstein condensate of $^{87}$Rb and $^{133}$Cs, \pra {\bf 84}, 011603 (2011).

\bibitem{hyperfine-sodium}
H. J. Miesner, D. M. Stamper-Kurn, J. Stenger, S. Inouye, A. P. Chikkatur, and W. Ketterle, 
Observation of metastable states in spinor Bose-Einstein condensates, \prl {\bf 82}, 2228 (1999).

\bibitem{Law-1997} 
C. K. Law, H. Pu, N. P. Bigelow, and J. H. Eberly,
Stability Signature in Two-Species Dilute Bose-Einstein Condensates, \prl {\bf 79}, 3105 (1997).

\bibitem{Chui-1998}
P. Ao and S. T. Chui, Binary Bose-Einstein condensate mixtures in weakly and strongly segregated phases, 
\pra {\bf 58}, 4836 (1998).

\bibitem{2002-Barankov-mix}
R. A. Barankov, Boundary of two mixed Bose-Einstein condensates, 
\pra {\bf 66}, 013612 (2002); B. D. Esry, Impact of spontaneous spatial symmetry breaking on the critical atom 
number for two-component Bose-Einstein condensates, \pra {\bf 58}, R3399(R) (1998).

\bibitem{2008-Abdu}
F.K. Abdullaev, A. Gammal, M. Salerno, and L. Tomio, 
Localized modes of binary mixtures of Bose-Einstein condensates in nonlinear optical lattices. 
\pra {\bf 77}, 023615 (2008).

\bibitem{2011-Pasquiou}
B. Pasquiou, E. Maréchal, G. Bismut, P. Pedri, L. Vernac, O. Gorceix, and B. Laburthe-Tolra, 
Spontaneous Demagnetization of a Dipolar Spinor Bose Gas in an Ultralow Magnetic Field, \prl {\bf 106}, 255303 (2011).

\bibitem{2012-Wilson}
R. M. Wilson, C. Ticknor, J. L. Bohn, and E. Timmermans, 
Roton immiscibility in a two-component dipolar Bose gas, 
\pra {\bf 86}, 033606 (2012).

\bibitem{phase-param}
L. Wen, W. M. Liu, Y. Cai, J. M. Zhang, and J. Hu, 
Controlling phase separation of a two-component Bose-Einstein condensate by confinement, 
\pra {\bf 85}, 043602 (2012).

\bibitem{pattinson} 
R. W. Pattinson, T. P. Billam, S. A. Gardiner, D. J. McCarron, H. W. Cho, S. L. Cornish, N. G. Parker, and N. P. Proukakis,
Equilibrium solutions for immiscible two-species Bose-Einstein condensates in perturbed harmonic traps,
\pra {\bf 87}, 013625 (2013);
R. W. Pattinson, {\it Two-component Bose-Einstein condensates: equilibria and dynamics at 
zero temperature and beyond}, PhD Thesis, Newcastle University,  Newcastle, UK, 2014.

\bibitem{salerno-2014} G. Filatrella, B.A. Malomed, and M. Salerno,
Domain walls and bubble droplets in immiscible binary Bose gases,  \pra {\bf 90}, 043629 (2014).

\bibitem{Yi2000} 
S. Yi and L. You, Trapped atomic condensates with anisotropic interactions,
Phys. Rev. A {\bf 61}, 041604 (R) (2000).

\bibitem{Santos2000} 
L. Santos, G. V. Shlyapnikov, P. Zoller, and M. Lewenstein,
Bose-Einstein Condensation in Trapped Dipolar Gases, 
Phys. Rev. Lett. {\bf 85}, 1791 (2000). Erratum in Phys. Rev. Lett.{\bf  88}, 139904 (2002).

\bibitem{2002-Goral}
K. G\'oral and L. Santos, Ground state and elementary excitations of single and binary Bose-Einstein condensates 
of trapped dipolar gases, \pra {\bf 66}, 023613 (2002).

\bibitem{2002-Giovanazzi} S. Giovanazzi, A. G\"orlitz, and T. Pfau, 
Tuning the Dipolar Interaction in Quantum Gases, Phys. Rev. Lett. {\bf 89}, 130401 (2002);
D. H. J. O'Dell, S. Giovanazzi, and C. Eberlein, 
Exact Hydrodynamics of a Trapped Dipolar Bose-Einstein Condensate,
Phys. Rev. Lett. {\bf 92}, 250401 (2004).

\bibitem{2005-Malomed} I. M. Merhasin, B.A. Malomed, and R. Driben, Transition to miscibility in a binary Bose-Einstein
condensate induced by linear coupling, J. Phys. B: At. Mol. Opt. Phys. {\bf 38}, 877-892 (2005).

\bibitem{2006-Bortolotti} D.C.E. Bortolotti, S. Ronen, J.L. Bohn, and D. Blume, 
Scattering Length Instability in Dipolar Bose-Einstein Condensates, \prl {\bf97}, 160402 (2006).
 
\bibitem{2007-Ronen}
S. Ronen, D.C.E. Bortolotti, and J.L. Bohn, Radial and angular rotons in trapped dipolar gases, \prl {\bf 98}, 
030406 (2007).

\bibitem{2008-Koch}
T. Koch, T. Lahaye, B. Fr\"ohlich, A. Griesmaier, and T. Pfau, Stabilization of a purely dipolar quantum gas against collapse, 
Nature Phys. {\bf 4}, 218-222 (2008).

\bibitem{2008-Lahaye}
T. Lahaye, J. Metz, B. Fröhlich, T. Koch, M. Meister, A. Griesmaier, T. Pfau, H. Saito, Y. Kawaguchi, and M. Ueda,
 d-Wave Collapse and Explosion of a Dipolar Bose-Einstein Condensate, \prl {\bf101}, 080401 (2008).

\bibitem{2008-Wilson}
R.M. Wilson, S. Ronen, J.L. Bohn, and H. Pu, Manifestations of the Ro- ton Mode in Dipolar Bose-Einstein Condensates,
\prl {\bf 100}, 245302 (2008).

\bibitem{Malomed-2010} 
G. Gligori\'c, A. Maluckov, M. Stepi\'c, L. Had\v{z}ievski, and B. Malomed, Transition to miscibility in linearly
coupled binary dipolar Bose-Einstein condensates, Phys. Rev. A {\bf 82}, 033624 (2010).

\bibitem{Zaman2011} 
M. Asad-uz-Zaman and D. Blume, Modification of roton instability due to the 
presence of a second dipolar Bose-Einstein condensate, Phys. Rev. A {\bf 83}, 033616 (2011).

\bibitem{martin-2012} A.D. Martin and P.B. Blakie, Stability and structure of an anisotropically trapped dipolar Bose-Einstein 
condensate: Angular and linear rotons, \pra {\bf 86}, 053623 (2012).

\bibitem{Adhikari-2012}
L. E. Young-S. and S. K. Adhikari, Mixing, demixing, and structure formation in a binary dipolar Bose-Einstein condensate, 
\pra  {\bf 86},  063611 (2012); Dipolar droplet bound in a trapped Bose-Einstein condensate, {\it ibid.} {\bf 87}, 013618 (2013).

\bibitem{2013-Bisset} R. N. Bisset and P. B. Blakie, Fingerprinting Rotons in a Dipolar Condensate: Super-Poissonian Peak in 
the Atom-Number Fluctuations, \prl  {\bf 110}, 265302 (2013);
R. Bisset, {\it Theoretical study of the trapped dipolar Bose gas in the ultra-cold regime}, 
PhD Thesis,  University of Otago, Dunedin, New Zeland,  2013.

\bibitem{2012-Aika-er}
K. Aikawa, A. Frisch, M. Mark, S. Baier, A. Rietzler, R. Grimm, and F. Ferlaino, Bose-Einstein condensation of Erbium, 
\prl {\bf 108}, 210401 (2012). For more details and references, see
A. Frisch, {\it Dipolar quantum gases of erbium}, PhD dissertation, University of Innsbruck, Innsbruck, 2014.

\bibitem{2008-Baranov}
M. A. Baranov, Theoretical progress in many-body physics with ultracold dipolar gases, Phys. Rep.~{\bf 464}, 
71-111 (2008).

\bibitem{2009-Lahaye} T. Lahaye, C. Menotti, L. Santos, M. Lewenstein, and T. Pfau, 
The physics of dipolar bosonic quantum gases, Rep. Prog. Phys.{\bf 72}, 126401 (2009).

\bibitem{2010-Ueda} M. Ueda, {\it Fundamentals and New Frontiers of Bose-Einstein Condensation},  
World Scientific Publ. Co., Singapore, 2010.

\bibitem{2005-Gries-cr} 
A. Griesmaier, J. Werner, S. Hensler, J. Stuhler, and T. Pfau, Bose-Einstein condensation of Chromium, 
\prl {\bf 94}, 160401 (2005);
T. Lahaye, T. Koch, B. Fr\"ohlich, M. Fattori, J. Metz, A. Griesmaier, S. Giovanazzi, and T. Pfau, 
Strong dipolar effects in a quantum ferrofluid, Nature {\bf 448}, 672?675 (2007).

\bibitem{2010-Ni}
K.-K. Ni, S. Ospelkaus, D. Wang, G. Qu\'em\'ener, B. Neyenhuis, M. H. G. de Miranda, J. L. Bohn, J. Ye, and D. S. Jin, 
Dipolar collisions of polar molecules in the quantum regime, Nature {\bf 464}, 1324?1328 (2010).

\bibitem{2010-Lu-Dy} 
M. Lu, S.H. Youn, and B.L. Lev, Trapping Ultracold Dysprosium: A Highly Magnetic Gas for Dipolar Physics, 
\prl {\bf 104}, 063001 (2010);
M. Lu, N. Q. Burdick, S. H. Youn, and B. L. Lev, Strongly dipolar Bose-Einstein condensate of Dysprosium, 
\prl {\bf 107}, 190401 (2011); 
Y. Tang, N. Q. Burdick, K. Baumann, B. L. Lev, Bose-€"Einstein condensation of $^{162}$Dy and $^{160}$Dy, 
New Journal of Physics {\bf 17}, 045006 (2015).

\bibitem{2016-Ferrier}
I. Ferrier-Barbut, H. Kadau, M. Schmitt, M. Wenzel, and T. Pfau, Observation of quantum droplets in a strongly dipolar Bose gas, 
\prl {\bf 116}, 215301 (2016).

\bibitem{1998-inouye} S. Inouye, M. R. Andrews, J. Stenger, H. J. Miesner, D. M. Stamper-Kurn, and W. Ketterle, 
Observation of Feshbach resonances in a Bose-Einstein condensate, Nature {\bf 392}, 151-154 (1998);
M.S. Hussein, E. Timmermans, P. Tommasini, and A.K. Kerman, Feshbach
resonances in atomic Bose-Einstein condensates, Phys. Rep. {\bf 315}, 199-230 (1999).

\bibitem{Gammal2006}
A. Gammal, T. Frederico, and L. Tomio, 
Critical number of atoms for attractive Bose-Einstein condensates with cylindrically symmetrical traps,
\pra {\bf 64}, 055602 (2001);
A. Gammal, L. Tomio, and T. Frederico, 
Critical numbers of attractive Bose-Einstein condensed atoms in asymmetric traps, \pra {\bf 66}, 043619 (2002);
M. Brtka, A. Gammal, and L. Tomio, 
Relaxation algorithm to hyperbolic states in Gross-Pitaevskii equation, Phys. Lett. A {\bf 359}, 339 (2006).

\bibitem{CPC1}
P. Muruganandam and S. K. Adhikari, 
Fortran programs for the time-dependent Gross-Pitaevskii equation in a fully anisotropic trap,
Comp. Phys. Commun. {\bf 180},  1888-1912 (2009); 
D. Vudragovi\'c, I. Vidanovi\'c, A. Bala\v{z}, P. Muruganandam and S. K. Adhikari, 
C programs for solving the time-dependent Gross-Pitaevskii equation in a fully anisotropic trap,
Comp. Phys. Commun. {\bf 183},  2021-2025 (2012).

\bibitem{CPC2}
R. K. Kumar, L. E. Young-S, D. Vudragovi\'{c}, A. Bala\v{z}, P. Muruganandam and S. K. Adhikari, 
Fortran and C programs for the time-dependent dipolar Gross-Pitaevskii equation in an anisotropic trap,
Comput. Phys. Commun. {\bf 195}, 117-128 (2015).

\bibitem{CUDA}
V. Lon\v{c}ar, A. Bala\v{z}, A. Bogojevi\'{c}, S. \v{S}krbi\'{c}, P. Muruganandam, S. K. Adhikari,
CUDA programs for solving the time-dependent dipolar Gross-Pitaevskii equation in an anisotropic trap,
Comp. Phys. Commun. {\bf 200}, 406-410 (2016).

\bibitem{efimov} 
V. Efimov, Energy levels arising from resonant two-body forces in a three-body system, 
Phys. Lett. B {\bf 33}, 563-564 (1970);
V. Efimov, Few-body physics: Giant trimers true to scale, Nature Phys. {\bf 5}, 533-534 (2009).

\bibitem{kraemer-2006}
T. Kraemer, {\it et al.}, Evidence for Efimov quantum states in an ultracold gas of caesium atoms,
Nature {\bf 440}, 315-318 (2006); F. Ferlaino and R. Grimm, Forty years of Efimov physics: How a 
bizarre prediction turned into a hot topic, Physics {\bf 3} (2010) 1-21.

\bibitem{3body}
T. Frederico, L. Tomio, A. Delfino, and A. E. A. Amorim, 
Scaling limit of weakly bound triatomic states, Phys. Rev. A {\bf 60} (1999) R9-R12;
E. Braaten, H.-W. Hammer, Universality in few-body systems with large scattering length, 
Phys. Rep. 428 (2006) 259-390. 

\bibitem{Kishor2016} R. K. Kumar, T. Sriraman, H. Fabrelli, P. Muruganandam, A. Gammal, 
Three-dimensional vortex structures in a rotating  dipolar Bose-Einstein condensate,
J. Phys. B: At. Mol. Opt. Phys. {\bf 49} 155301 (2016).

\end{thebibliography}
\end{document}